\documentclass[
 aip,
 amsmath,amssymb,
 reprint,
]{revtex4-1}

\usepackage{epsfig}
\usepackage{dcolumn}
\usepackage{bm}
\usepackage[utf8]{inputenc}
\usepackage[T1]{fontenc}
\usepackage{mathptmx}
\usepackage{xcolor}
\usepackage{layouts}
\usepackage{graphicx}
\usepackage{caption}
\usepackage{subcaption}
\usepackage{tabularray}
\usepackage{comment}
\usepackage[hmargin=1cm]{geometry}
\usepackage{ulem}
\usepackage{amsmath}
\usepackage{lipsum}
\usepackage{multirow}
\usepackage{physics}
\usepackage{braket}

\definecolor{auburn}{rgb}{0.43, 0.21, 0.1}

\newcommand{\C}[1]{{\color{red}{Q: #1}}}
\definecolor{zgreen}{rgb}{0.0, 0.26, 0.15}
\newcommand{\quasi}[1]{{\it quasi}}

\def\F7{C$_{70}$~}
\def\C80{C$_{80}$}
\def\Sc3N{Sc$_{3}$N}
\def\Y3N{Y$_{3}$N}

\begin{document}
\preprint{UTEP/005}

\title{How well do one-electron self-interaction-correction methods perform for systems with fractional electrons?}

\author{ Rajendra R. Zope}
\email{rzope@utep.edu}
\author{Yoh Yamamoto}
\author{Tunna Baruah}
\affiliation{Department of Physics, The University of Texas at El Paso, El Paso, TX 79968, USA }
\date{\today }

\begin{abstract}
 Recently developed locally scaled self-interaction correction (LSIC) is a one-electron SIC method 
 that, when used with a ratio of kinetic energy densities ($z_{\sigma}$) as iso-orbital indicator, performs remarkably well for 
 both thermochemical properties as well as for barrier heights overcoming the paradoxical behavior of
the well-known Perdew-Zunger self-interaction correction (PZSIC) method.
 In this work, we examine how well the LSIC method performs for the delocalization error.
 Our results show that both LSIC and PZSIC methods correctly describe the dissociation of H$_2^{+}$ and He$_2^{+}$ but LSIC is 
 overall more accurate than the PZSIC method.
 Likewise,  in the case of the vertical ionization energy of an ensemble of isolated He atoms, the LSIC and PZSIC methods do not exhibit delocalization errors. 
 For the fractional charges, both LSIC and PZSIC  significantly reduce the deviation from linearity in the energy versus number of electrons curve, with PZSIC performing superior for C, Ne, and Ar atoms while  for Kr they perform similarly. 
 The LSIC performs well at the endpoints (integer occupations) while substantially
 reducing the deviation. The dissociation of LiF shows both LSIC and PZSIC dissociate into 
 neutral Li and F but only LSIC exhibits charge transfer from Li$^+$ to F$^-$ at the expected distance
 from the experimental data and accurate ab initio data. Overall both the PZSIC and LSIC methods 
 reduce the delocalization errors substantially.
\end{abstract}

\maketitle

\section {Introduction}
The Kohn-Sham formulation of the density functional theory (KS-DFT)\cite{PhysRev.136.B864,PhysRev.140.A1133} has a major share in the application of electronic structure studies in computational chemistry and  
in solid state physics. The KS-DFT, in principle, allows exact calculations of the ground state energy and the electron density. Practical applications, however,  require approximations to the exchange-correlation component of the total energy functional. One major reason for the ubiquitous usage of 
the KS-DFT for computation of the ground state total energies, total energy differences, equilibrium structures, etc. 
is due to the development of accurate approximations to exchange-correlation functionals over the last three decades. 
These approximations differ in complexity, computational cost, and accuracy  and have been broadly categorized by Perdew and Schmidt as various “rungs” on “Jacob’s ladder” to the heaven of chemical accuracy.\cite{perdew2001jacob} 
The rungs on this ladder begin with the simple local spin density approximation (LSDA) followed by the semi-local functionals like the generalized gradient approximation (GGA) and the meta-GGA functionals. The fourth rung corresponds to hyper-GGA functionals that include the exact exchange, while the fifth rung is a random phase approximation.  Designing accurate exchange-correlation functionals is an active area of research and a large number of functionals have been developed. The applications of the DFT with these functionals have provided a wealth of information about the properties of molecules, nanomaterials, and solids. 

Despite the widespread usage, most density functional approximations (DFAs) suffer from 
self-interaction error (SIE). The SIE  arises from the fact that self-Coulomb energy is 
not exactly canceled by the approximate self-exchange-correlation energy. 
In a system with a single electron, the electron's only interaction is with a nucleus.
Consequently, the exchange-correlation energy $E_{xc}[\rho]$ should 
cancel the Coulomb energy  $U[\rho]$ in a single electron system. As the correlation energy $E_c[\rho]$ is zero for the 
single electron case, the $E_x[\rho] $ must equal $U[\rho]$. This condition is not satisfied
by most DFAs. The difference $U[\rho]-E_{xc}$ is the SIE 
made by 
the functional and is known as one-electron SIE.  
This error is also responsible for poor asymptotic description of the 
exchange potentials of the DFAs causing significant errors in the eigenvalue of 
the electron.  For example, the widely used PBE GGA functional underestimates 
the eigenvalue of the hydrogen atom by 
 6 eV.  A similar underestimation of the first ionization potential (IP) 
 (as an absolute of the highest occupied  orbital (HOO) energy)
 is seen in the many-electron systems.
The one-electron SIE 
in a many-electron system can be removed by subtracting the SIE of each orbital.
One-electron self-interaction correction (SIC) approaches, of which Perdew-Zunger SIC (PZSIC) is the most well-known, eliminate SIE on an orbital-by-orbital basis 
as follows
\begin{equation}\label{eq:pzsic}
    E^{DFA-SIC}= E^{DFA}[\rho_\uparrow,\rho_\downarrow]-\sum_{i\sigma}^{occ}\left( U[\rho_{i\sigma}]+E_{XC}^{DFA}[\rho_{i\sigma},0] \right).
\end{equation}
Here, $\rho_i$ is the density of the $i^{th}$ orbital.
Such an approach was proposed by Lindgren fifty years ago\cite{lindgren1971statistical} (in 1971) to augment Hartree theory to include 
self-interaction-free local exchange using Slater's X$\alpha$ exchange functional 
which was subsequently extended by Perdew\cite{perdew1979orbital} and coworkers\cite{perdew1981self,zunger1980self} to include the correlation.
A nice detailed overview of the early SIC methods and their applications
is given by the Perdew-Zunger in their  well-cited 1981 article.\cite{perdew1981self}

A number of problems, such as unbound atomic anions, incorrect asymptotic behavior of the potential, overestimation of polarizabilities of conjugated systems, excessive electron delocalization, etc., can be corrected with some success using the one-electron SIC methods. However, the dependence on the orbital density complicates the implementation of SIC methods and requires the use of localized orbitals for size 
consistency.\cite{heaton1983self,pederson1984local}
These issues and the paradoxical performance of the PZSIC method\cite{perdew2015paradox}
that dominate the one-electron SIC methods have prevented the one-electron SIC methods from becoming mainstream electronic structure methods. Instead, DFAs that include Hartree-Fock (HF) exchange using various schemes\cite{schmidt2016one} are widely used to mitigate the erroneous behavior of local and semi-local DFAs due to the SIE.

  The many-electron SIE and localization/delocalization errors are related concepts that 
  pertain to the DFA errors of the fractionally charged fragments of a system.
  Perdew and coworkers\cite{perdew1982density} %developed exact DFT
  discussed the exact constraint in DFT for fractional particles,  which as noted by Perdew in Ref. 
  \onlinecite{ruzsinszky2007density}
  was motivated by his early work on the development of one-electron SIC (PZSIC) method.
  They showed that the density functional total energy of an $N$-electron system should vary linearly between integer electron numbers and should exhibit derivative discontinuities at integer electron numbers. 
  At integer electron the slope of $E$ versus $N$ curve on the electron-deficient side is 
  equal to the negative of the exact vertical IP and on the electron-rich 
  side is the electron affinity of the integer system.
  The total energy of an $N$-electron system, on the other hand, evolves as a convex curve when the electronic charge varies between $N$ and $N+1$ electrons when using approximate density functionals.
    DFAs artificially lower the energy of a fractional electron system, resulting in a convex curve rather than a linear curve between integer electron values.  
This convex behavior of DFAs that is a deviation from expected linearity is known as delocalization error.\cite{zheng2012delocalization}
Unlike DFAs, the HF approximation exhibits concave behavior which is referred to as localization 
error.
The terms delocalization/localization errors are often used to describe SIEs 
in DFAs for 
many-electron systems.\cite{
zheng2012delocalization,dwyer2011dispersion,li2018localized,johnson2013extreme}
Sometimes they are also called the {\it many-electron} SIE in literature.\cite{ruzsinszky2006spurious} 
Designing functionals or methods that minimize, if not eliminate, the effect of SIE 
(or delocalization error) is one of the most significant unsolved problems in practical 
DFT development.
Aside from the one-electron SIC methods that remove SIE in an orbitalwise fashion, 
several other approaches have been proposed for eliminating or reducing SIE, like von Weizs\"acker kinetic energy density-based SICs (an exchange functional by Becke and Roussel derived from Taylor expansion of exchange hole),\cite{becke1989exchange,becke1983hartree} regional SIC,\cite{tsuneda2003regional,tsuneda2014self}  local hybrid functional,\cite{jaramillo2003local,kaupp2007local,schmidt2014self} long-range asymptotic corrections,\cite{latter1955atomic} Koopmans-compliant functionals,\cite{dabo2014piecewise,borghi2014koopmans} atomic SIC (ASIC),\cite{pemmaraju2007atomic} multiconfiguration pair-DFT,\cite{li2014multiconfiguration} and 
localized orbital scaling correction (LOSC).\cite{li2018localized,su2020preserving} and the
rung-3.5 approach\cite{janesko2021replacing}.
More recently, machine and deep learning have been used to mitigate or eliminate the delocalization 
errors.\cite{nagai2020completing,kirkpatrick2021pushing,bogojeski2020quantum,dick2021highly} 

We have recently introduced a locally scaled SIC method (LSIC) that
has shown promising results with the simplest local spin density functional.\cite{zope2019step,yamamoto2023self} 
 LSIC  uses a pointwise iso-orbital indicator to identify the one-electron self-interaction regions in the many-electron system (see Section~\ref{sec:theory}) and to determine the magnitude of SIC in the many-electron regions. 
 The LSIC method is not only exact for the one-electron densities but it
 also recovers the uniform electron gas limit of the uncorrected DFA, 
 and reduces to the well-known
 PZSIC 
 method as a special case.
In  particular,
it is free from the paradoxical behavior of the well known PZSIC method, by giving 
simultaneous good results for thermochemistry and barrier heights.
LSIC  provides improved performance with respect to the standard PZSIC method for a wide range of electronic structure 
properties.\cite{zope2019step,waterpolarizability,akter2021static, doi:10.1021/acs.jpca.1c10354,mishra2022study,akter2021well,yamamoto2023self,romero2023spin}
 Early applications of the LSIC method were performed either in a perturbative fashion 
using the PZSIC densities or in a quasi-self-consistent manner. Very recently, we have implemented  self-consistent 
 LSIC method 
 for the case where the ratio of Weizs\"acker and Kohn–Sham kinetic energy densities is used as an iso-orbital indicator.\cite{yamamoto2023self}
This enabled
us to examine how well it performs in cases where delocalization errors are known to be pronounced,
which is the topic of the present manuscript. For this purpose, we use Fermi-L\"owdin SIC (FLOSIC) approach\cite{pederson2014communication}
in which Fermi-L\"owdin orbitals (FLOs)\cite{luken1982localized,luken1984localized} are used as localized orbitals. For comparison, we also  perform 
calculations using the PZSIC method within the FLOSIC approach.\cite{pederson2014communication,PhysRevA.103.042811} In the following section, we briefly describe 
the FLOSIC approach and the LSIC method.

\section{Theory and Computational methods}\label{sec:theory}
    The total energy in the LSIC method is given by
\begin{align}\label{eq:lsic}
    E^{LSIC}= E^{DFA}[\rho_\uparrow,\rho_\downarrow]-\sum_{i\sigma}^{occ}\left( U^{LSIC}[\rho_{i\sigma}]+E_{XC}^{LSIC}[\rho_{i\sigma},0] \right)
\end{align}
where 
\begin{align}
    & U^{LSIC}[\rho_{i\sigma}] = \frac 1 2 \int d^3{r} \,\, z_\sigma(\mathbf{r}) \,\, \rho_{i\sigma}(\mathbf{r}) \,\, \int d^3{r'} \frac{\rho_{i\sigma}(\mathbf{r'})}{|\mathbf{r}-\mathbf{r'}|}  \\
    & E_{XC}^{LSIC}[\rho_{i\sigma},0]=\int d^3{r}  \,\,z_\sigma (\mathbf{r}) \,\, \varepsilon_{XC}^{DFA}([\rho_{i\sigma}],\mathbf{r}).
\end{align}
The summation $i$ in Eq.~(\ref{eq:lsic}) runs over all occupied orbitals of spin $\sigma$. $z_{\sigma}$ is a suitable iso-orbital indicator 
that identifies the single electron regions. In this work, we choose 
the ratio of   the von Weizs\"acker  $\tau_\sigma^W$  
and total positive-definite kinetic energy densities $\tau_\sigma$ 
of spin $\sigma$, that is, 
$$z_{\sigma}(\mathbf{r})=\frac{\tau_\sigma^W(\mathbf{r})}{\tau_\sigma(\mathbf{r})}.$$
The iso-orbital indicator, $z_\sigma$  varies between $0$ and $1$, with $0$ corresponding to the uniform electron density limit where
non-empirical local and semilocal DFAs are exact by construction, and $1$ to the one-electron density limit 
where the full PZSIC correction yields exact results. 
Thus LSIC with chosen $z_\sigma$ is exact in both the  
one-electron density limit as well as the uniform electron density limit. 
Alternative choices for iso-orbital indicators are possible\cite{romero2021local,doi:10.1063/5.0041646}
but we use $z_\sigma$ as defined above as it has 
performed well for various properties.\cite{zope2019step,waterpolarizability,akter2021static, doi:10.1021/acs.jpca.1c10354,mishra2022study,akter2021well,yamamoto2023self,romero2023spin}
The PZSIC method is also exact 
(by construction) in the one-electron density limit but not in the uniform density limit. LSIC reduces to PZSIC 
for a special case of $z_\sigma = 1 $.
The size consistency requires the use of localized orbital densities $\rho_i$  for evaluating orbitalwise SIC  in Eq. (\ref{eq:lsic}).

In the FLOSIC approach of Pederson, Ruzsinsky, and Perdew,\cite{pederson2014communication} 
these localized orbital densities are obtained as the orthogonalized Fermi 
orbitals.\cite{luken1982localized,leonard1982quadratically,luken1984localized}
This work and early  FLOSIC applications used  FLOSIC interchangeably with PZSIC, but the FLOSIC 
approach can also be used in other variants of one-electron SIC methods.\cite{lundin2001novel,vydrov2006scaling,yamamoto2020improvements,zope2019step,romero2023complexity}
The FLOs are 
derived from the Kohn-Sham orbitals $\psi_{j\sigma}$ as
\begin{equation}\label{FerOrb}
 \phi_{i\sigma}^\mathrm{FO}(\mathbf{r}) 
  = \frac{{\gamma}_\sigma(\mathbf{a}_{i\sigma},\mathbf{r})} {\sqrt{\rho_\sigma(\mathbf{a}_{i\sigma})}}.
 \end{equation}
Here, ${\gamma}_\sigma = \sum_{j}^{occ} f_j \, \psi_{j\sigma}^*(\mathbf{r})  \psi_{j\sigma}(\mathbf{a}_{i\sigma}) $, 
is the single-particle Kohn-Sham density matrix, $f_j$ is the occupation number of Kohn-Sham orbital $\psi_j$, and $\mathbf{a}_{i\sigma}$ are points in real space called Fermi orbital descriptors or FODs. 
The FLOSIC method is unitarily invariant as any set of orbitals spanning the same occupied space leads to the same Fermi orbitals (FO).  
Since the FO are normalized but not orthonormal, they are orthogonalized using 
the L\"owdin symmetric orthogonalization scheme,\cite{lowdin1950non,luken1982localized} resulting in the orthonormal FLOs that 
are then used to evaluate the SIC energy (Eq.~(\ref{eq:pzsic})). 
The computational cost of PZSIC and LSIC  methods within the FLOSIC 
formalism is discussed in earlier studies.\cite{yang2017full,zope2019step,yamamoto2023self}

Since different FOD positions yield different FLOs, and therefore different total energies,
minimizing the total energy in FLOSIC implies finding optimal FOD positions. 
Gradients of the SIC energy with respect to the FODs\cite{pederson2015fermi,pederson2015self} 
are then used to optimize the FOD positions. 
Some PZSIC and LSIC calculations reported in this work are performed perturbatively 
using the LSDA densiy. In such calculations 
the FODs were optimized by freezing the LSDA electron density.

PZSIC and LSIC calculations in this work are performed using the development version of the FLOSIC code based 
on the UTEP-NRLMOL code.\cite{FLOSICcode,FLOSICcodep} 
We have considered LSIC with LSDA functional as 
it performs well with the LSDA compared to the PZSIC method.
LSIC with GGAs and meta-GGAs functionals performs only slightly better than the PZSIC, possibly
due to the gauge problems.\cite{bhattarai2020step} The gauge problem occurs as exchange-correlation energy 
densities of these approximations are not in the Hartree gauge. 
A gauge transformation is required to correct the energy density for GGAs and meta-GGAs.\cite{bhattarai2020step,PhysRevA.77.012509}
On the other hand, as shown in Ref.~\onlinecite{yamamoto2023self}, removing SIE using a properly developed SIC method can eliminate the majority of LSDA errors and predict atomization energies and barrier heights accurately.

We used the LSDA correlation functional parameterized as PW92\cite{PhysRevB.46.6671} and the NRLMOL basis set that is triple zeta quality.\cite{porezag1999optimization} 
For the numerical integration, an adaptive mesh is used.\cite{pederson1990variational}  
In the  FLOSIC scheme self-consistency can be obtained either
with the Krieger-Li-Iafrate approximation\cite{PhysRevA.103.042811}  or using 
the Jacobi update approach.\cite{yang2017full} We have compared both approaches and found that they give very similar results for properties
considered in this work. The results presented are obtained with the Jacobi update approach.\cite{yang2017full}
A tolerance of 10$^{-6}$ $E_h$ is used as the SCF convergence criteria, and 
force criteria of 10$^{-3}$ $E_h/a_0$ is used for the FOD optimizations.

We use the following terminology to describe the results obtained in various method: LSIC-LSDA for self-consistent LSIC results with the LSDA functional and PZSIC-LSDA for self-consistent PZSIC results with the LSDA functionals.
We denote the perturbative energy evaluation of LSIC-LSDA on PZSIC-LSDA density by LSIC-LSDA@PZSIC-LSDA.
All other perturbative calculations are denoted in the same manner. 
These perturbative evaluations are used to gauge the effect of energy correction and density correction of a given functional separately. 
As explained in Ref.~\onlinecite{akter2021static} LSIC can be applied to scale SIC potentials instead 
of the energy density. This  approach is referred to as quasi-SCF LSIC or qLSIC.

\section{Results and discussion}

\begin{figure}
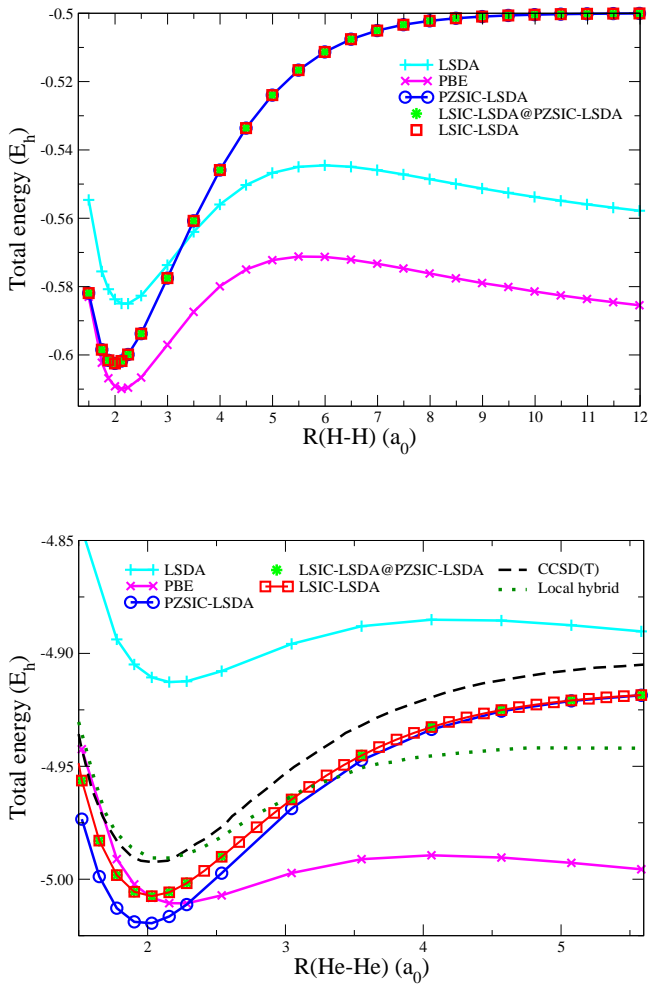

\includegraphics[width=0.9\columnwidth]{Fig1a.eps}\\
\vspace{1cm}
\includegraphics[width=0.9\columnwidth]{Fig1b.eps}
\caption{\label{fig:h2p_he2p} Dissociation curves for  H$_2^+$ (top) and  He$_2^+$ (bottom) with
various methods. 
The dissociation curve with perturbative LSIC  with PZSIC density (denoted by LSIC-LSDA@PZSIC-LSDA) and self-consistent LSIC
are essentially identical.
The CCSD(T)/cc-pVQZ  curve 
and local hybrid curve for He$_2^+$ are from Ref.~\onlinecite{ruzsinszky2007density} and Ref.~\onlinecite{schmidt2014self},
respectively. 
}
\end{figure}

\subsection{Dissociation curves of  H$_2^+$ and He$_2^+$}

  The dissociation curves of the H$_2^+$ and He$_2^+$ molecules are often used to highlight the effect 
of delocalization errors (fractional electron behavior).\cite{zhang1998challenge,ruzsinszky2005binding,bryenton2023delocalization,bao2018self} 
Figure 1  shows the energy of the H$_2^+$ and He$_2^+$ ions as a function of interatomic distance $R$. 
H$_2^+$ is the simplest molecule with one electron yet most DFAs make significant errors in the dissociation limit.
For example, PBE and LSDA errors are about 2-2.5 eV with LSDA having a smaller error than PBE. Both, the PZSIC 
and LSIC methods being exact for one-electron densities exhibit the correct dissociation limit.
The dissociation curve of the He$_2^+$ is also shown in  Fig.~\ref{fig:h2p_he2p}.  This 3-electron system 
is a representative of the many-electron system. The dissociation curve for the LSDA, PBE,
PZSIC-LSDA, LSIC-LSDA@PZSIC-LSDA, and LSIC-LSDA are compared
with the coupled-cluster singles, doubles, and perturbative triples [CCSD(T)] benchmark results from Ref.~\onlinecite{ruzsinszky2007density}.  
The LSDA significantly underestimates energy for all bond distances. 
The PZSIC-LSDA improves over LSDA results but overestimates the energy, especially
near equilibrium. On the other hand, LSIC-LSDA provides a significantly better description near
equilibrium by scaling down the correction and transitions towards the PZSIC-LSDA curve as the 
interatomic distance increases. It coincides with PZSIC-LSDA in the dissociation limit.
The LSIC-LSDA curve is parallel to the CCSD(T) with almost a constant shift of 0.4 eV.
The local hybrid (from Ref. \onlinecite{schmidt2014self}), which 
is an alternative approach to remove SIE 
by locally 
mixing HF exchange with DFA,\cite{jaramillo2003local,maier2019local} performs better near the equilibrium but significantly
underestimates the energies at the dissociation limit.
The LSIC-LSDA@PZSIC-LSDA and the LSIC-LSDA plots are on top of each 
other, illustrating that the difference between the PZSIC and LSIC results is primarily 
due to the difference in the PZSIC and LSIC energy functionals. In Section \ref{sec:sie4x4-set}, we discuss the 
dissociation of symmetric cations of other elements.
The dissociation of symmetric cations within the PZSIC 
has been studied previously.\cite{ruzsinszky2007density} Perdew and coworkers have examined the role of 
SIE in the molecular dissociation limits using various functionals with  and without
SIC. They used the  PZSIC method and scaled-down PZSIC wherein the SIC energy is scaled 
by orbital-dependent scalar constants.\cite{vydrov2006scaling}
They observed that PZSIC applied to the PBE functional produces
S-like curves for the electron number dependence of the ground state 
energies of atoms, with the largest curvatures near integer electron numbers.\cite{vydrov2007tests}

\subsection{Delocalization error in the ionization energy of an ensemble of He atoms}
 To illustrate the delocalization errors of the DFAs, Yang and coworkers have extended 
 the example of He$_2^+$ in dissociation limit by including additional widely separated
 He atoms.
 We follow this representative system  chosen to study in
 earlier studies of delocalization errors\cite{doi:10.1021/acs.jpclett.7b02705}
 and thus consider 
 a model system of an ensemble of He atoms in a ring configuration
 where the He 
 atoms are placed 10 \AA\, apart. This is a many-electron
 system.
\begin{figure*}
    \centering
    \begin{subfigure}[t]{0.35\textwidth}
        \centering
        \includegraphics[width=\linewidth]{Fig2a.eps}
        \label{HeRing_IP}
    \end{subfigure}
    \hfill
    \begin{subfigure}[t]{0.35\textwidth}
        \centering
        \includegraphics[width=\linewidth]{Fig2b.eps}
        \label{HeRing_LSIC}
    \end{subfigure}
    \hfill
    \begin{subfigure}[t]{0.25\textwidth}
    \includegraphics[width=\linewidth]{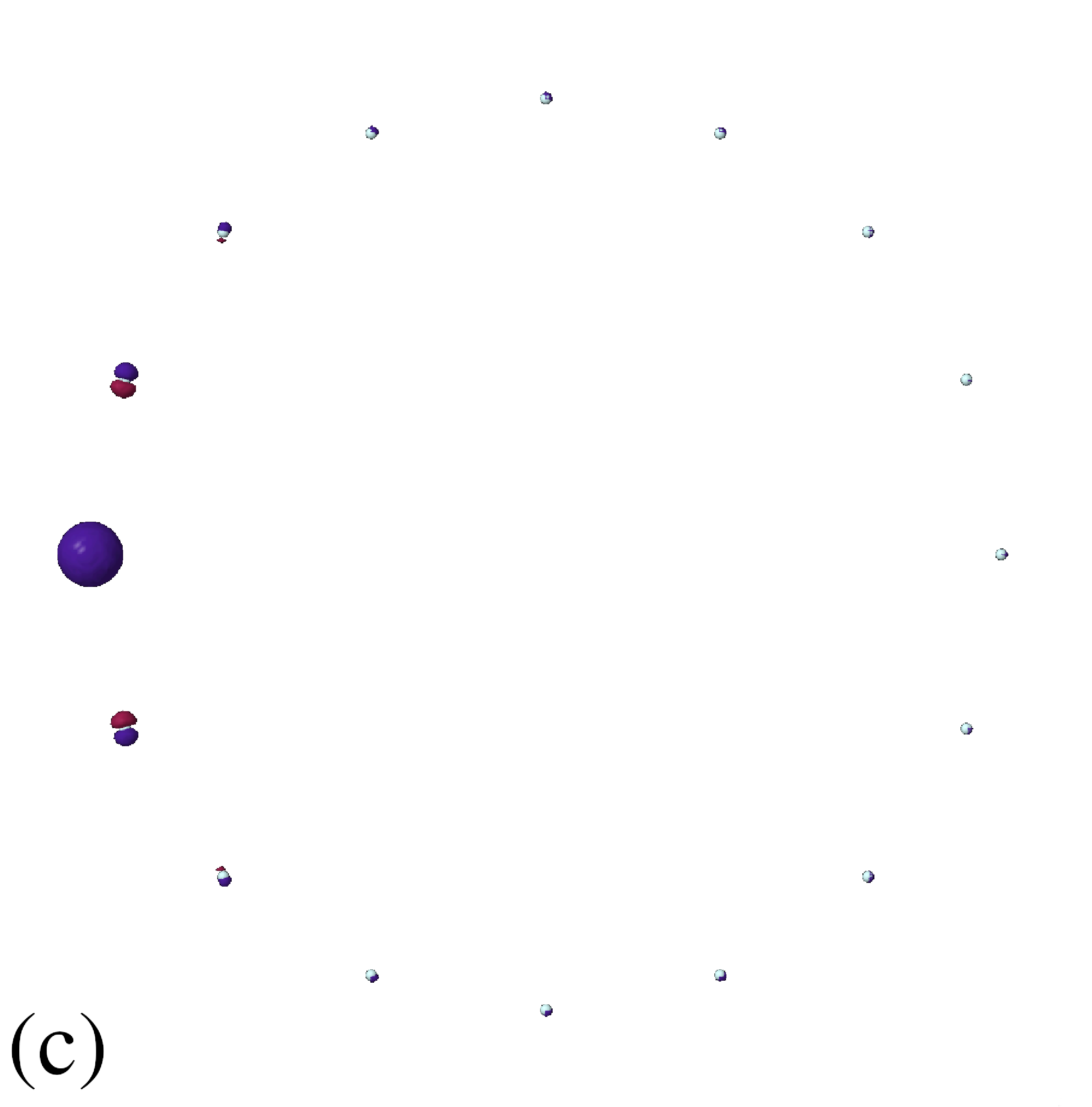}
    \label{Hedensity}
    \end{subfigure}    

    \caption{The ionization energy of ring of He atoms as a function of a number of atoms using various methods. In the LSIC case (middle figure),  0.9 
    electron was removed due to self-consistent convergence problems. Note that while in the LSDA, PBE, BLYP, and SCAN the ionization energy decreases as atoms 
    are added due to delocalization error, it remains constant in PZSIC and LSIC indicating that these methods are free from delocalization
    error. Fig. 1(c) shows the density difference of He$_{16}$ ring in neutral and cationic states. The spin density is localized within the  PZSIC method. }
    \label{fig:herium_ring}
\end{figure*}
 The 10 \AA\, separation between the atoms is large enough
 to remove any interaction between them. The calculation of IP
 with DFA for such non-interacting atoms nicely portrays
 the delocalization problem. The IP for a chain 
 of non-interacting atoms should be the same as that of a 
single atom.  Within the DFAs (LSDA, GGA, and many hybrids)
the IP of such a system decreases as more atoms are added 
to the chain.\cite{cohen2012challenges,doi:10.1021/acs.jpclett.7b02705} 
This decrease in IP is evident in Fig. \ref{fig:herium_ring}(a)  
 which shows the  IP as a function of a number of 
atoms for the PBE functional. When an electron is removed 
from the He ring,  it leaves the system with a hole that is
delocalized over the ring. 
The spin density of  the cation of a ring of He atoms 
integrates to one electron, which within the GGA (PBE)
is spread over the whole ring.  This systematic increase in delocalization as the number of He atoms are added 
 diminishes the IP within DFAs (LSDA, PBE, BLYP, and SCAN studied in this work).
 Correcting for SIE localizes the extra electron density to mainly
over a single He atom as shown in Fig. \ref{fig:herium_ring}(c). 
 Inspection of the FLOSIC local orbitals shows that they are mainly localized around each atom.
The effect of localization is also seen from the ionization energy vs. the number of atom plots with FLOSIC (cf. Fig.~
\ref{fig:herium_ring}(a)).
Within PZSIC-LSDA, the IP versus
 number of helium atom curve is essentially a straight line with zero slope.  The IP of a chain
 remains close to that of a single atom.  Although this particular system 
 had not been studied using the PZSIC scheme earlier, the present result 
 is in conformity with earlier works that the PZSIC method significantly reduces the many-electron SIE of DFAs.\cite{vydrov2007tests}
 We note in passing that the error made by LSDA is due to energy lowering for the
fractional electron(s). Enforcing the integer electron occupation within the LSDA
does give the correct behavior of the constant IP as a function of He atoms.
It would be interesting to see how LSIC works in this situation. We were unable to obtain convergence for the whole one-electron removal using the LSIC approach for certain sizes. As a result, 0.9 electron was removed from the non-interacting ensemble of He atoms
in the ring configuration.
This is sufficient to know the density delocalization trend within the LSIC method. The results
are depicted in Fig.~\ref{fig:herium_ring}(b). 
Like the PZSIC method, the IP remains unchanged for a  chain of increasing numbers of
non-interacting He atoms suggesting that the LSIC method does not suffer from delocalization error,
at least for the chosen many-electron representative system.

\begin{figure*}
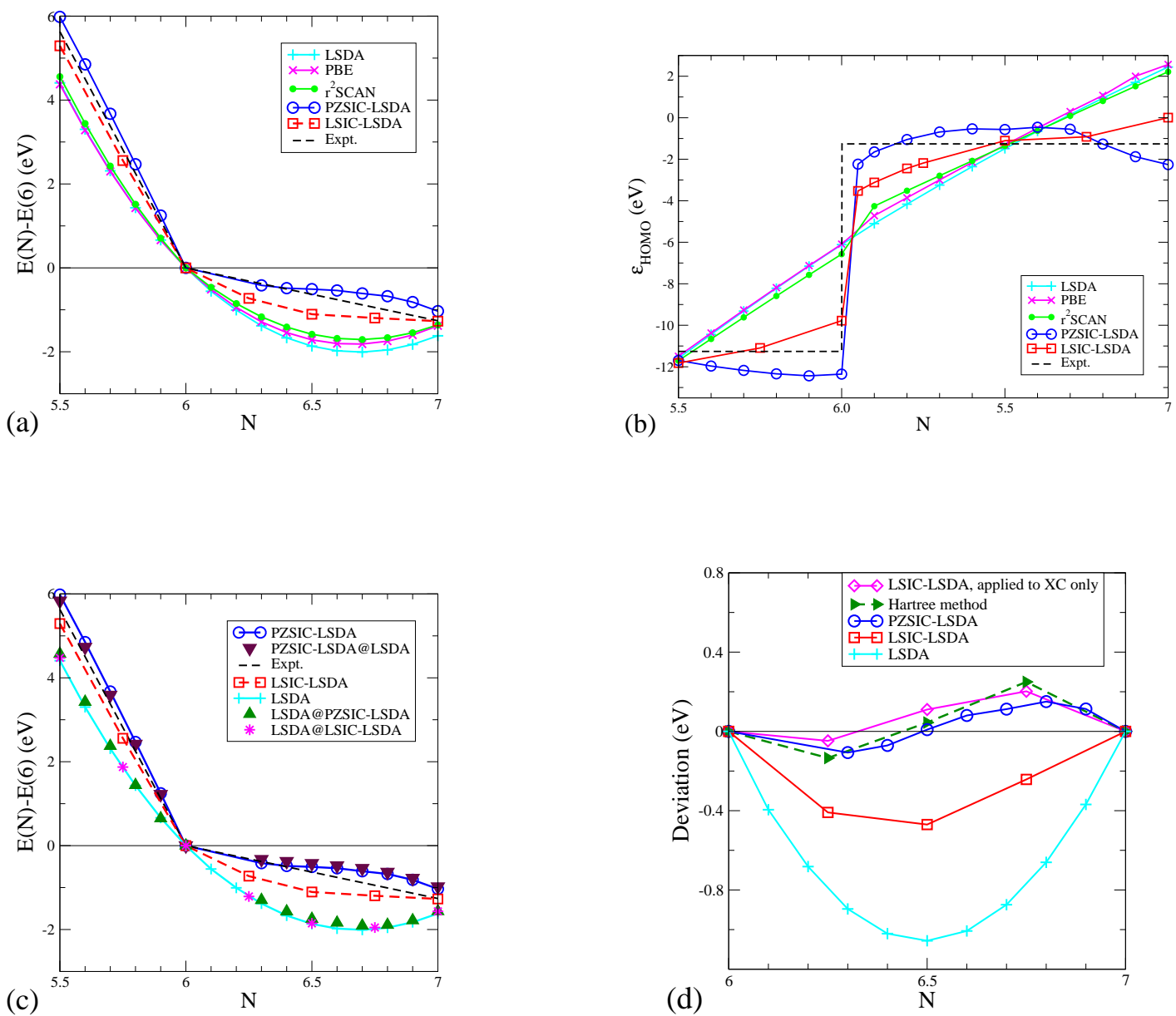

    \centering
    \begin{subfigure}[t]{0.45\textwidth}
        \centering
        \includegraphics[width=0.8\linewidth]{Fig3a.eps}
    \end{subfigure}
    \hfill
    \begin{subfigure}[t]{0.45\textwidth}
        \centering
        \includegraphics[width=\linewidth]{Fig3b.eps}
    \end{subfigure}
 
    \vspace{2cm}
    
    \begin{subfigure}[t]{0.45\textwidth}
        \centering
        \includegraphics[width=0.8\linewidth]{Fig3c.eps}
    \end{subfigure}    
    \hfill
    \begin{subfigure}[t]{0.45\textwidth}
        \centering
        \includegraphics[width=0.85\linewidth]{Fig3d.eps}
    \end{subfigure}    
      \caption{
      (a) %and (c)
      Total energy (in eV) of C atom with respect to the respective neutral atom energy E(6) 
      as a function of number of electrons in various approximations. 
      (b) The eigenvalues of the highest occupied orbital (in eV). The dashed lines are IP and EA from the experiment. 
      The derivative discontinuity is seen for the two SIC methods whereas LSDA and PBE do not have any discontinuity.
      (c) The DFA (SIC) total energy evaluated with the SIC (DFA) density relative to E(6). 
      (d) Deviation from linearity (in eV) between the electron numbers $N=6$ to 7. 
      LSIC applied to exchange-correlation (XC) only gives smaller deviations than the LSIC applied to both Coulomb and XC. Similarly, the Hartree method that only corrects for self-Coulomb energies gives smaller deviations.
      }\label{subfig:carbon-totalenergy}
    \label{fig:carbonatom}    
\end{figure*}

\subsection{Fractional charge behavior of atoms}

To obtain further information about the behavior of the LSIC method for fractional 
charges, we compute the energy of C atom as a function of 
varying number of electrons $N$ from C$^{+}$ ($N=5.5$)  to C$^{-}$ ($N=7$). 
Supplemental 
basis functions were added for an adequate description of carbon anion. 
The generalized Janak's theorem states that the orbital energy of a fractional charged state is linear as a function of charge.\cite{janak1978proof, cohen2008fractional}
The results with LSIC, PZSIC, and DFAs are shown in Figure \ref{fig:carbonatom}(a). 
It is clear from the figure that LSDA, PBE, and r$^2$SCAN lack derivative 
discontinuity at $N=6$ while it is conspicuous in the case of LSIC and PZSIC methods.
Both LSIC and PZSIC curves show smaller deviations from linearity compared to the LSDA, PBE and r$^2$SCAN functionals.
Between the PZSIC and LSIC, the PZSIC has a smaller deviation from linearity 
while\ LSIC is closer to the exact results near 
endpoints. The PZSIC and LSIC deviations are opposite to each other.
In Fig. \ref{fig:carbonatom} (b) we present the 
HOO eigenvalue 
of carbon as the electron number varies from 5.5 to 7.0.  We chose HOO eigenvalue
because of the numerical evidence that 
the HOO eigenvalue closely mimics the ionization energy in the case of atoms and 
molecules and vertical detachment energy in the case of anions.\cite{vargas2020importance,ufondu2023vertical}
The exact curve is deduced from experimental IP and EA and shown as a dotted line.
The LSDA and PBE
as expected do not show any discontinuity. A slight change in the slope is seen 
in the case of r$^2$SCAN functional. Both the PZSIC and LSIC eigenvalues 
exhibit derivative discontinuity as electron number changes through 
the integer value of $6$. The LSIC method exhibits smaller derivative 
discontinuity compared to the PZSIC method, this is consistent with 
total energy change as a function of the electron shown in Fig. \ref{fig:carbonatom}(a).

To understand the role of one-electron SIC methods in reducing the delocalization error, we 
   analyze the SIC contribution to the energy for the fractional charges.
   The DFA 
   part of the SIC energy expression is plotted in Fig. \ref{fig:carbonatom}(c) indicated with 
   LSDA@PZSIC-LSDA and LSDA@LSIC-LSDA. This is the DFA total energy ($E^{DFA}$ in Eq. (\ref{eq:pzsic}))
   evaluated with the SIC density. As seen in the figure, these curves are essentially identical to 
   that obtained from the DFA (LSDA in this case), thus using SIC density has no effect 
   on the delocalization (fractional electron) error in this case.  This, in turn, 
   suggests that a simple perturbative scheme wherein SIC energy is evaluated perturbatively
   using DFA density may be sufficient in such a case. The calculations show that this is 
   indeed the case.
   As can be seen in Fig. \ref{fig:carbonatom}(c), the  
   PZSIC-LSDA@LSDA  (PZSIC evaluated perturbatively using LSDA orbitals)
   practically overlaps with the self-consistent PZSIC-LSDA curve. 
   As shown in the subsequent section, this approach appears to work for
   molecules (as opposed to single center C atom) where bond 
   distances are slightly stretched.
   Fig. \ref{fig:carbonatom}(d) also shows 
   the LSIC curve wherein the LSIC is applied only to the exchange-correlation term
   i.e.,
   full Hartree correction (self-Coulomb energy) is used while the scaling the self-exchange-correlation energy density as 
   $E^{DFA}[\rho_\uparrow,\rho_\downarrow]-\sum_{i\sigma}^{occ}\left( U[\rho_{i\sigma}]+E_{XC}^{LSIC}[\rho_{i\sigma},0] \right)$.
   This 
   results in a reduction of the fractional electron error by a similar magnitude to
   the PZSIC method. 
   The contribution to the correction from the exchange-correlation term  is about 
   the same  magnitude in all three SIC schemes. 
   Thus, for minimal SIE in the fractional electron systems, full SIC to the Coulomb energy
   is desirable, that is,  the one-electron SIC 
   the scheme should be designed like Hartree plus SIC (scaled or unscaled) 
   exchange-correlation energy.\cite{lindgren1971statistical}
   Such a scaled SIC scheme may work better with GGA or meta-GGA since the SIC contributions
   from the exchange-correlation terms in these functional is often greater than the Coulomb 
   contribution, resulting thereby in positive SIC corrections.
   It may be difficult, however, to obtain good energies at integer values.

\subsection{Dissociation of LiF molecule}

\begin{figure*}
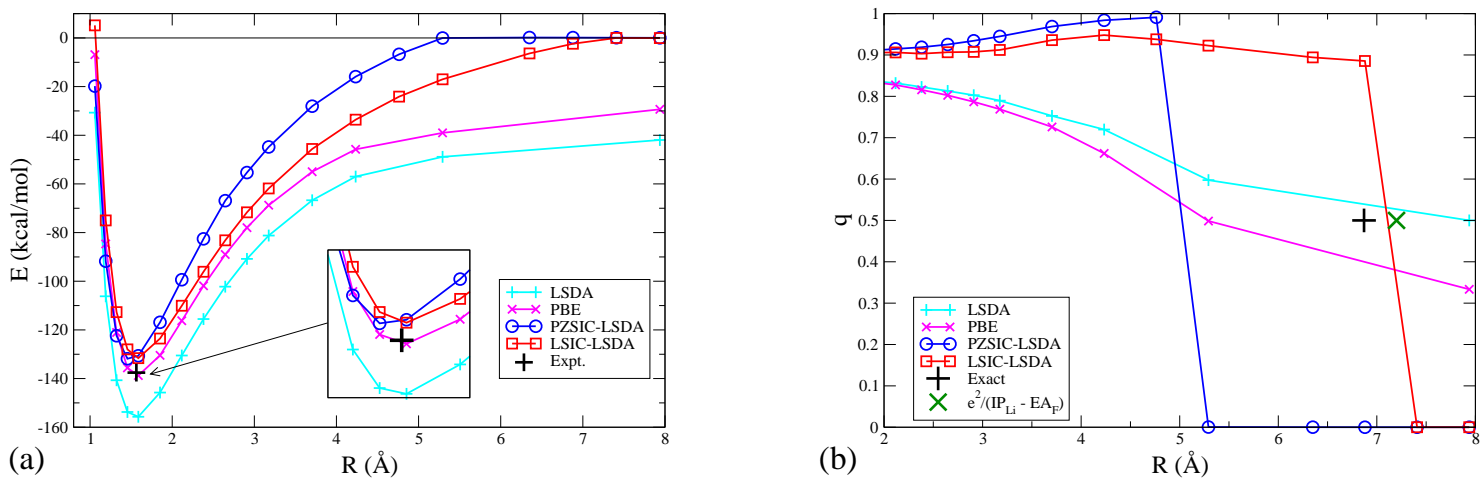

    \centering
    \begin{subfigure}[t]{0.45\textwidth}
        \centering
        \includegraphics[width=\linewidth]{Fig4a.eps}
        \label{subfig:LiF_dissociation} 
    \end{subfigure}
    \hfill
        \begin{subfigure}[t]{0.45\textwidth}
        \centering
        \includegraphics[width=\linewidth]{Fig4b.eps}
        \label{subfig:LiF-Mulliken} 
    \end{subfigure}
    \caption{Dissociation behavior of the LiF molecule:
    (a) Dissociation energy (in kcal/mol) 
    and
    (b) Mulliken population charge on lithium atom as a function of bond length. 
        Both PZSIC and LSIC have abrupt charge transfer; only LSIC shows accurate critical distance.
    }\label{fig:LiF} 
\end{figure*}

   We now consider a dissociation of heteroatomic LiF molecule. Many DFAs predict the dissociation of LiF into Li and F atoms with 
   {\it spurious} fractional charge.  The energy required to transfer an electron from Li atom 
   to the F atom to form Li$^{+}$ and  F$^{-}$ ions pair is given by the difference in the 
   IP 
   of Li atom and the electron affinity (EA) of the F atom plus the 
   electrostatic interaction between the two ions. By considering  the two ions as point 
   charges that are separated by a distance of $R$, one can estimate the charge transfer energy $E_{CT}$ as 
   \begin{equation}
    E_{CT} =  (IP)_{Li} - (EA)_{F} - \frac{1}{R}.
   \end{equation}
    The point charge assumption is only valid for a reasonably large separation.
    The experimental IP for the Li atom is 5.39 eV and the EA of the F atom is 3.4 eV.\cite{johnson2013nist} Since
     $(IP)_\text{Li} \, >\,  (EA)_\text{F}$  the LiF molecule will dissociate 
     into neutral Li and  F atoms. In the vicinity of the equilibrium bond distance, the 
     Li and F are in the charge states. By setting $E_{CT}$ to zero one obtains 
     the critical distance $R_c = ((IP)_\text{Li} - (EA)_\text{F})^{-1}$, at which the electrostatic attraction of the ions
     equals the energy difference between neutral atoms and ions. Using 
     the experimental values of IP and EA one expects $R_c$ to be 7.2 \AA.
    In addition, Ref.~\onlinecite{li2018density} reports the critical distance $R_c$ to be $6.87$, which includes the non-adiabatic effect.
    The dissociation curve of  LiF is shown in Fig. \ref{fig:LiF}(a). 
    Similar to the PBE and LSDA, LSIC-LSDA gives accurate equilibrium bond length.
    Like the PZSIC-LSDA, the LSIC-LSDA also dissociates correctly into
    the neutral Li and F atoms.
    Figure \ref{fig:LiF}(b) shows the Mulliken charge on the Li atom as a function of 
    bond distance.
  As the interatomic separation increases, a charge transfer, Li$^+$+F$^- \rightarrow$ Li + F, should occur 
   at $R_c$. 
   The charge transfer occurs abruptly both in the PZSIC and LSIC methods. 
   As noted by Vydrov and coworkers 
   this charge transfer occurs at $R_c$ of 3.5 \AA \, for the HF method,\cite{vydrov2007tests} roughly
   half of the experimental estimate of 7.2 \AA. Their PZSIC calculations 
   with PBE functional similarly predict the transition to neutral dissociation products at a much 
   smaller distance of  $3.5$ \AA.  On the other hand, our PZSIC-LSDA calculations provide a significantly
   better prediction of  $R_c$ ($\sim$5~\AA) compared to the result ($3.5$ \AA) of Vydrov and coworkers.
   The difference in two PZSIC $R_c$ values may be due to the functional rather than the differences in the SIC
   implementations.
   On the other hand, in the LSIC method, the transition to 
   neutral fragments occurs around $7$ \AA, in excellent agreement with the experimental 
   estimate of $7.2$ \AA. 
   Thus, LSIC outperforms PZSIC both near equilibrium and at the dissociation limit for the LiF
   but exhibits a smaller derivative discontinuity than the PZSIC.

\begin{table}
    \centering
    \caption{The midpoint error $2E(X^{+0.5})- E(X) - E(X^{+}) $ in kcal/mol (for $X=$ Ne, Ar, and Kr) 
     within LSDA, PBE, PZSIC-LSDA and LSIC-LSDA methods.
     Since the maximum error of the PZSIC method may not occur at the midpoint, their maximum error
     is also reported in a separate column (shown in bold font for those occurring at non midpoint).
     }
    \begin{ruledtabular}     
    \begin{tabular}{cccc}
    Atom & Method & Midpoint error & Max error \\
    \hline
       Ne  &  LSDA & -102.5 & -102.5  \\
       Ne  &  PBE  & -97.6   & -97.6\\
       Ne  & PZSIC-LSDA & 3.4 & {\bf 17.2} \\
       Ne  & LSIC-LSDA  & -42.7 &  -42.7 \\
    \hline
       Ar  &  LSDA & -67.6 & -67.6 \\
       Ar  &  PBE  & -62.7  & -62.7 \\
       Ar  & PZSIC-LSDA & -1.1 & {\bf 9.4} \\
       Ar  & LSIC-LSDA & -31.7 & -31.7 \\
    \hline
       Kr  & LSDA & -58.5 & -58.5 \\
       Kr  &  PBE & -55.4  & -55.4\\
       Kr  & PZSIC-LSDA & 18.6 & 18.6 \\
       Kr  & LSIC-LSDA & -22.0 & -22.0 \\
    \end{tabular}
    \end{ruledtabular}
    \label{tab:my_label}
\end{table}

\subsection{Midpoint errors in Ne, Ar, and Kr}

     We have also tried to generate dissociation curves for the symmetric charged 
     radicals  Ne$_2^{+}$, Ar$_2^{+}$, and  Kr$_2^{+}$ but were unable to obtain the 
     energies for all bond separations due to difficulties in obtaining self-consistent 
     convergence. So in order to obtain a qualitative understanding of energy lowering 
     due to fractional charges, we compute  the midpoint error  $2E(X^{+0.5})- E(X) - E(X^{+}) $ 
     for $X=$ Ne, Ar, and Kr. Most DFAs  have convex behavior (
      $2E(X^{+0.5}) <  E(X) + E(X^{+}) $) and exhibit delocalizing behavior. 
     On the other hand,  the HF approximation has  $2E(X^{+0.5}) > E(X) - E(X^{+}) $ 
     and has a localizing tendency.
     The computed midpoint values are shown in Table \ref{tab:my_label}. 
     As expected midpoint error is large for LSDA and PBE and is substantially
     reduced within LSIC-LSDA. For atoms Ne and Ar, the midpoint error is 
     the smallest for the PZSIC. However, our results show that the PZSIC total energies show 
     somewhat different behavior compared to DFA or LSIC-LSDA. For C, Ne, and Ar atoms PZSIC shows concave
     behavior from the midpoint to the left and convex behavior
     in the other half of energy versus electron curves. That is, 
     the maximum deviation from linearity for these systems in the case of the PZSIC 
     method does not occur at the midpoint. We therefore report  
     the maximum errors in a separate column for 
     the PZSIC method.
     On the other hand, 
     for Kr,  PZSIC exhibits convex behavior indicating a localizing tendency.
     For Kr, LSIC-LSDA midpoint error is opposite in sign but comparable in
     magnitude to the PZSIC.

\begin{table}
\caption{\label{tab:sie4x4}MAE of SIE4x4 (in kcal/mol).}
\footnotetext{Reference \onlinecite{yamamoto2019fermi}}
\begin{ruledtabular}
\begin{tabular}{lc}
 Method & MAE(kcal/mol) \\ \hline
LSDA$^\text{a}$	& 27.5\\
\hline
PZSIC-LSDA$^\text{a}$	& 3.0\\ 
PZSIC-LSDA@LSDA &  3.5  \\
\hline
LSIC-LSDA@PZSIC-LSDA	& 2.6 \\
qLSIC-LSDA    & 3.8 \\
LSIC-LSDA          & 4.0 \\
LSIC-LSDA@LSDA &  4.8\\
\end{tabular}
\end{ruledtabular}
\end{table}

\subsection{SIE4x4}\label{sec:sie4x4-set}

The SIE4x4 set was introduced to examine the SIE which is more pronounced when the bonds
are stretched.\cite{C7CP04913G} This set consists 
of dissociation energies of set of four positively charged dimer systems with four distances $R$
between the fragments where $R/R_e =$ 1.0, 1.25, 1.5, and 1.75 with $R_e$ is the equilibrium distance.
Self-consistent calculations of these charged dimers in a stretched configuration are difficult
due to convergence issues.
Typically, in most SIC calculations we use a simple (linear) mixing 
scheme with a mixing parameter value of $\alpha=0.15$.
This means 85\% of the original (previous iteration) Hamiltonian or density matrix with 15\% 
of current iteration to construct an input Hamiltonian or density matrix.
We found that in the case of stretched dimers,
higher values of $\alpha$ are needed to obtain self-consistent convergence. 
We used $\alpha$ in the range 0.2-0.4  for (NH$_3$)$_2^+$ and (H$_2$O)$_2^+$. 
The calculated dissociation energies are summarized in Table~\ref{tab:sie4x4}.
The one-electron SIE-free PZSIC performs well in these cases
with the mean absolute error 
(MAE) of PZSIC-LSDA to be 2.8 kcal/mol.  The LSIC-LSDA@PZSIC-LSDA (perturbative LSIC-LSDA using the 
PZSIC density) performs similarly well  (MAE = 2.6 kcal/mol). 
In both SCF and qLSIC, 
MAEs are slightly larger than PZSIC and perturbative LSIC.
The MAEs are 4.0 and 3.8 kcal/mol for SCF and qLSIC, respectively.
As discussed in an earlier section,  in the case of the C atom with fractional charges, the energy versus the number of electrons curve obtained with the PZSIC method using the LSDA density gave nearly identical results to those obtained from self-consistent PZSIC calculations.
Unlike the carbon atoms, molecules in the SIE4x4 set are two centered 
and therefore more prone to delocalization error.
The results however show that the MAE of PZSIC-LSDA@LSDA is 3.5 kcal/mol 
comparable to that of self-consistent PZSIC (3.0 kcal/mol) calculations.
This again indicates that the major improvement from LSDA to PZSIC-LSDA is from the energy 
functional while the difference in density does not affect much. 
The SIE4x4 set considers intermolecular distance only up to 1.75 R$_{eq}$, where the density correction with SIC may not be important.

\section{Summary and Outlook}
Our previous studies\cite{zope2019step,waterpolarizability,akter2021static, doi:10.1021/acs.jpca.1c10354,mishra2022study,akter2021well,yamamoto2023self,romero2023spin}
have shown that the LSIC method performs well 
both for the thermochemical properties as well as the properties that
involve energetics at stretched bonds such as barrier heights, thus 
offering a resolution to the long-standing paradoxical nature of 
the most known one-electron SIC method called the PZSIC method.
The performance of the PZSIC for the fractionally charged systems 
have been studied in the past.\cite{vydrov2004effect,ruzsinszky2006spurious,ruzsinszky2007density,https://doi.org/10.1002/jcc.26168}
It has been found that the PZSIC 
method significantly reduces the errors of the DFAs in such cases,
but PZSIC predictions of energy at integer occupations are not 
sufficiently accurate.\cite{vydrov2006scaling,zope2019step}
The LSIC performs well for the integer occupations giving 
good performance for thermochemical and related 
properties\cite{zope2019step,akter2021static,ufondu2023vertical,mishra2022study,yamamoto2023self} 
and as found in this work significantly reduces the delocalization 
error. The vertical ionization energy of an ensemble of 
well separated He atoms remain constants both in LSIC and PZSIC 
as atoms are added suggesting that these methods  are
free from delocalization error. The energy versus electron 
number studied for different atoms shows that LSIC reduces
the delocalization error significantly, but it appears that 
using LSIC with full Hartree correction can further improve the performance
for fractional charges albeit at the expense of accuracy at 
the integer values. It is therefore desirable to design 
the iso-orbital indicator such that good accuracy of LSIC
at integer electron number is maintained while the error 
for the fractional charge is further diminished. We considered
LiF as a prototype system to study the dissociation of heteronuclear
diatomic molecules.
The LiF 
dissociation is correctly described by both the LSIC and 
PZSIC methods but LSIC bond length is closer to the 
experimental value while PZSIC underestimates it.
 Similarly,  in LSIC the conversion from 
 charged to neutral separated species occurs
at a distance of 7 {\AA}  in excellent agreement with 
experimental estimate of 7.2 {\AA}  while in the PZSIC 
it occurs at a  much shorter ( 5 \AA) distance. Overall 
performance of LSIC shows that the LSIC reduces the delocalization 
error substantially. 
One interesting finding of the present work is that a
simpler approach wherein the SIC 
is obtained perturbatively using DFA orbitals can also 
be equally effective as the self-consistent 
PZSIC method in  some cases (where density corrections are small)
such as SIE4x4 dataset or energies of atoms with fractional
number of electrons.

\section*{Supplementary Material} 
See supplementary material for the total energy and deviation from linearity plots using PBE and the SIC methods for the neon and argon atoms.

\section*{Data Availability Statement}
The data that supports findings in this article is provided within the manuscript and supplementary information.

\section*{Acknowledgment}
This work was supported by the US Department of Energy, Office of Science, Office of Basic Energy Sciences, as part of the Computational Chemical Sciences Program under Award No. DE-SC0018331. 
Support for computational time at the Texas Advanced  Computing Center (TACC),  and National Energy Research Scientific Computing Center (NERSC) is gratefully acknowledged.

\bibliography{master}

\begin{thebibliography}{84}%
\makeatletter
\providecommand \@ifxundefined [1]{%
 \@ifx{#1\undefined}
}%
\providecommand \@ifnum [1]{%
 \ifnum #1\expandafter \@firstoftwo
 \else \expandafter \@secondoftwo
 \fi
}%
\providecommand \@ifx [1]{%
 \ifx #1\expandafter \@firstoftwo
 \else \expandafter \@secondoftwo
 \fi
}%
\providecommand \natexlab [1]{#1}%
\providecommand \enquote  [1]{``#1''}%
\providecommand \bibnamefont  [1]{#1}%
\providecommand \bibfnamefont [1]{#1}%
\providecommand \citenamefont [1]{#1}%
\providecommand \href@noop [0]{\@secondoftwo}%
\providecommand \href [0]{\begingroup \@sanitize@url \@href}%
\providecommand \@href[1]{\@@startlink{#1}\@@href}%
\providecommand \@@href[1]{\endgroup#1\@@endlink}%
\providecommand \@sanitize@url [0]{\catcode `\\12\catcode `\$12\catcode
  `\&12\catcode `\#12\catcode `\^12\catcode `\_12\catcode `\%12\relax}%
\providecommand \@@startlink[1]{}%
\providecommand \@@endlink[0]{}%
\providecommand \url  [0]{\begingroup\@sanitize@url \@url }%
\providecommand \@url [1]{\endgroup\@href {#1}{\urlprefix }}%
\providecommand \urlprefix  [0]{URL }%
\providecommand \Eprint [0]{\href }%
\providecommand \doibase [0]{http://dx.doi.org/}%
\providecommand \selectlanguage [0]{\@gobble}%
\providecommand \bibinfo  [0]{\@secondoftwo}%
\providecommand \bibfield  [0]{\@secondoftwo}%
\providecommand \translation [1]{[#1]}%
\providecommand \BibitemOpen [0]{}%
\providecommand \bibitemStop [0]{}%
\providecommand \bibitemNoStop [0]{.\EOS\space}%
\providecommand \EOS [0]{\spacefactor3000\relax}%
\providecommand \BibitemShut  [1]{\csname bibitem#1\endcsname}%
\let\auto@bib@innerbib\@empty
%</preamble>
\bibitem [{\citenamefont {Hohenberg}\ and\ \citenamefont
  {Kohn}(1964)}]{PhysRev.136.B864}%
  \BibitemOpen
  \bibfield  {author} {\bibinfo {author} {\bibfnamefont {P.}~\bibnamefont
  {Hohenberg}}\ and\ \bibinfo {author} {\bibfnamefont {W.}~\bibnamefont
  {Kohn}},\ }\bibfield  {title} {\enquote {\bibinfo {title} {Inhomogeneous
  electron gas},}\ }\href {\doibase 10.1103/PhysRev.136.B864} {\bibfield
  {journal} {\bibinfo  {journal} {Phys. Rev.}\ }\textbf {\bibinfo {volume}
  {136}},\ \bibinfo {pages} {B864--B871} (\bibinfo {year} {1964})}\BibitemShut
  {NoStop}%
\bibitem [{\citenamefont {Kohn}\ and\ \citenamefont
  {Sham}(1965)}]{PhysRev.140.A1133}%
  \BibitemOpen
  \bibfield  {author} {\bibinfo {author} {\bibfnamefont {W.}~\bibnamefont
  {Kohn}}\ and\ \bibinfo {author} {\bibfnamefont {L.}~\bibnamefont {Sham}},\
  }\bibfield  {title} {\enquote {\bibinfo {title} {Self-consistent equations
  including exchange and correlation effects},}\ }\href {\doibase
  10.1103/PhysRev.140.A1133} {\bibfield  {journal} {\bibinfo  {journal} {Phys.
  Rev.}\ }\textbf {\bibinfo {volume} {140}},\ \bibinfo {pages} {A1133--A1138}
  (\bibinfo {year} {1965})}\BibitemShut {NoStop}%
\bibitem [{\citenamefont {Perdew}\ and\ \citenamefont
  {Schmidt}(2001)}]{perdew2001jacob}%
  \BibitemOpen
  \bibfield  {author} {\bibinfo {author} {\bibfnamefont {J.~P.}\ \bibnamefont
  {Perdew}}\ and\ \bibinfo {author} {\bibfnamefont {K.}~\bibnamefont
  {Schmidt}},\ }\bibfield  {title} {\enquote {\bibinfo {title} {Jacob's ladder
  of density functional approximations for the exchange-correlation energy},}\
  }in\ \href {\doibase 10.1063/1.1390175} {\emph {\bibinfo {booktitle} {AIP
  Conference Proceedings}}},\ Vol.\ \bibinfo {volume} {577}\ (\bibinfo
  {organization} {American Institute of Physics},\ \bibinfo {year} {2001})\
  pp.\ \bibinfo {pages} {1--20},\ \Eprint
  {http://arxiv.org/abs/https://aip.scitation.org/doi/pdf/10.1063/1.1390175}
  {https://aip.scitation.org/doi/pdf/10.1063/1.1390175} \BibitemShut {NoStop}%
\bibitem [{\citenamefont {Lindgren}(1971)}]{lindgren1971statistical}%
  \BibitemOpen
  \bibfield  {author} {\bibinfo {author} {\bibfnamefont {I.}~\bibnamefont
  {Lindgren}},\ }\bibfield  {title} {\enquote {\bibinfo {title} {A statistical
  exchange approximation for localized electrons},}\ }\href@noop {} {\bibfield
  {journal} {\bibinfo  {journal} {Int. J. Quantum Chem.}\ }\textbf {\bibinfo
  {volume} {5}},\ \bibinfo {pages} {411--420} (\bibinfo {year}
  {1971})}\BibitemShut {NoStop}%
\bibitem [{\citenamefont {Perdew}(1979)}]{perdew1979orbital}%
  \BibitemOpen
  \bibfield  {author} {\bibinfo {author} {\bibfnamefont {J.}~\bibnamefont
  {Perdew}},\ }\bibfield  {title} {\enquote {\bibinfo {title} {Orbital
  functional for exchange and correlation: self-interaction correction to the
  local density approximation},}\ }\href@noop {} {\bibfield  {journal}
  {\bibinfo  {journal} {Chem. Phys. Lett.}\ }\textbf {\bibinfo {volume} {64}},\
  \bibinfo {pages} {127--130} (\bibinfo {year} {1979})}\BibitemShut {NoStop}%
\bibitem [{\citenamefont {Perdew}\ and\ \citenamefont
  {Zunger}(1981)}]{perdew1981self}%
  \BibitemOpen
  \bibfield  {author} {\bibinfo {author} {\bibfnamefont {J.~P.}\ \bibnamefont
  {Perdew}}\ and\ \bibinfo {author} {\bibfnamefont {A.}~\bibnamefont
  {Zunger}},\ }\bibfield  {title} {\enquote {\bibinfo {title} {Self-interaction
  correction to density-functional approximations for many-electron systems},}\
  }\href {\doibase 10.1103/PhysRevB.23.5048} {\bibfield  {journal} {\bibinfo
  {journal} {Phys. Rev. B}\ }\textbf {\bibinfo {volume} {23}},\ \bibinfo
  {pages} {5048} (\bibinfo {year} {1981})}\BibitemShut {NoStop}%
\bibitem [{\citenamefont {Zunger}, \citenamefont {Perdew},\ and\ \citenamefont
  {Oliver}(1980)}]{zunger1980self}%
  \BibitemOpen
  \bibfield  {author} {\bibinfo {author} {\bibfnamefont {A.}~\bibnamefont
  {Zunger}}, \bibinfo {author} {\bibfnamefont {J.}~\bibnamefont {Perdew}}, \
  and\ \bibinfo {author} {\bibfnamefont {G.}~\bibnamefont {Oliver}},\
  }\bibfield  {title} {\enquote {\bibinfo {title} {A self-interaction corrected
  approach to many-electron systems: Beyond the local spin density
  approximation},}\ }\href@noop {} {\bibfield  {journal} {\bibinfo  {journal}
  {Solid State Commun.}\ }\textbf {\bibinfo {volume} {34}},\ \bibinfo {pages}
  {933--936} (\bibinfo {year} {1980})}\BibitemShut {NoStop}%
\bibitem [{\citenamefont {Heaton}, \citenamefont {Harrison},\ and\
  \citenamefont {Lin}(1983)}]{heaton1983self}%
  \BibitemOpen
  \bibfield  {author} {\bibinfo {author} {\bibfnamefont {R.~A.}\ \bibnamefont
  {Heaton}}, \bibinfo {author} {\bibfnamefont {J.~G.}\ \bibnamefont
  {Harrison}}, \ and\ \bibinfo {author} {\bibfnamefont {C.~C.}\ \bibnamefont
  {Lin}},\ }\bibfield  {title} {\enquote {\bibinfo {title} {Self-interaction
  correction for density-functional theory of electronic energy bands of
  solids},}\ }\href@noop {} {\bibfield  {journal} {\bibinfo  {journal} {Phys.
  Rev. B}\ }\textbf {\bibinfo {volume} {28}},\ \bibinfo {pages} {5992}
  (\bibinfo {year} {1983})}\BibitemShut {NoStop}%
\bibitem [{\citenamefont {Pederson}, \citenamefont {Heaton},\ and\
  \citenamefont {Lin}(1984)}]{pederson1984local}%
  \BibitemOpen
  \bibfield  {author} {\bibinfo {author} {\bibfnamefont {M.~R.}\ \bibnamefont
  {Pederson}}, \bibinfo {author} {\bibfnamefont {R.~A.}\ \bibnamefont
  {Heaton}}, \ and\ \bibinfo {author} {\bibfnamefont {C.~C.}\ \bibnamefont
  {Lin}},\ }\bibfield  {title} {\enquote {\bibinfo {title} {Local-density
  {H}artree-{F}ock theory of electronic states of molecules with
  self-interaction correction},}\ }\href {\doibase 10.1063/1.446959} {\bibfield
   {journal} {\bibinfo  {journal} {J. Chem. Phys.}\ }\textbf {\bibinfo {volume}
  {80}},\ \bibinfo {pages} {1972--1975} (\bibinfo {year} {1984})},\ \Eprint
  {http://arxiv.org/abs/https://doi.org/10.1063/1.446959}
  {https://doi.org/10.1063/1.446959} \BibitemShut {NoStop}%
\bibitem [{\citenamefont {Perdew}\ \emph {et~al.}(2015)\citenamefont {Perdew},
  \citenamefont {Ruzsinszky}, \citenamefont {Sun},\ and\ \citenamefont
  {Pederson}}]{perdew2015paradox}%
  \BibitemOpen
  \bibfield  {author} {\bibinfo {author} {\bibfnamefont {J.~P.}\ \bibnamefont
  {Perdew}}, \bibinfo {author} {\bibfnamefont {A.}~\bibnamefont {Ruzsinszky}},
  \bibinfo {author} {\bibfnamefont {J.}~\bibnamefont {Sun}}, \ and\ \bibinfo
  {author} {\bibfnamefont {M.~R.}\ \bibnamefont {Pederson}},\ }\bibfield
  {title} {\enquote {\bibinfo {title} {Paradox of self-interaction correction:
  How can anything so right be so wrong?}}\ }in\ \href {\doibase
  10.1016/bs.aamop.2015.06.004} {\emph {\bibinfo {booktitle} {Adv. At. Mol.
  Opt. Phys.}}},\ Vol.~\bibinfo {volume} {64}\ (\bibinfo  {publisher}
  {Elsevier},\ \bibinfo {year} {2015})\ pp.\ \bibinfo {pages}
  {1--14}\BibitemShut {NoStop}%
\bibitem [{\citenamefont {Schmidt}\ and\ \citenamefont
  {K{\"u}mmel}(2016)}]{schmidt2016one}%
  \BibitemOpen
  \bibfield  {author} {\bibinfo {author} {\bibfnamefont {T.}~\bibnamefont
  {Schmidt}}\ and\ \bibinfo {author} {\bibfnamefont {S.}~\bibnamefont
  {K{\"u}mmel}},\ }\bibfield  {title} {\enquote {\bibinfo {title} {One-and
  many-electron self-interaction error in local and global hybrid
  functionals},}\ }\href@noop {} {\bibfield  {journal} {\bibinfo  {journal}
  {Phys. Rev. B}\ }\textbf {\bibinfo {volume} {93}},\ \bibinfo {pages} {165120}
  (\bibinfo {year} {2016})}\BibitemShut {NoStop}%
\bibitem [{\citenamefont {Perdew}\ \emph {et~al.}(1982)\citenamefont {Perdew},
  \citenamefont {Parr}, \citenamefont {Levy},\ and\ \citenamefont {{Balduz
  Jr}}}]{perdew1982density}%
  \BibitemOpen
  \bibfield  {author} {\bibinfo {author} {\bibfnamefont {J.~P.}\ \bibnamefont
  {Perdew}}, \bibinfo {author} {\bibfnamefont {R.~G.}\ \bibnamefont {Parr}},
  \bibinfo {author} {\bibfnamefont {M.}~\bibnamefont {Levy}}, \ and\ \bibinfo
  {author} {\bibfnamefont {J.~L.}\ \bibnamefont {{Balduz Jr}}},\ }\bibfield
  {title} {\enquote {\bibinfo {title} {Density-functional theory for fractional
  particle number: derivative discontinuities of the energy},}\ }\href@noop {}
  {\bibfield  {journal} {\bibinfo  {journal} {Phys. Rev. Lett.}\ }\textbf
  {\bibinfo {volume} {49}},\ \bibinfo {pages} {1691} (\bibinfo {year}
  {1982})}\BibitemShut {NoStop}%
\bibitem [{\citenamefont {Ruzsinszky}\ \emph {et~al.}(2007)\citenamefont
  {Ruzsinszky}, \citenamefont {Perdew}, \citenamefont {Csonka}, \citenamefont
  {Vydrov},\ and\ \citenamefont {Scuseria}}]{ruzsinszky2007density}%
  \BibitemOpen
  \bibfield  {author} {\bibinfo {author} {\bibfnamefont {A.}~\bibnamefont
  {Ruzsinszky}}, \bibinfo {author} {\bibfnamefont {J.~P.}\ \bibnamefont
  {Perdew}}, \bibinfo {author} {\bibfnamefont {G.~I.}\ \bibnamefont {Csonka}},
  \bibinfo {author} {\bibfnamefont {O.~A.}\ \bibnamefont {Vydrov}}, \ and\
  \bibinfo {author} {\bibfnamefont {G.~E.}\ \bibnamefont {Scuseria}},\
  }\bibfield  {title} {\enquote {\bibinfo {title} {Density functionals that are
  one-and two-are not always many-electron self-interaction-free, as shown for
  {H}$_2^+$, {He}$_2^+$, {LiH}$^+$, and {Ne}$_2^+$},}\ }\href@noop {}
  {\bibfield  {journal} {\bibinfo  {journal} {J. Chem. Phys.}\ }\textbf
  {\bibinfo {volume} {126}} (\bibinfo {year} {2007})}\BibitemShut {NoStop}%
\bibitem [{\citenamefont {Zheng}\ \emph {et~al.}(2012)\citenamefont {Zheng},
  \citenamefont {Liu}, \citenamefont {Johnson}, \citenamefont
  {Contreras-Garc{\'\i}a},\ and\ \citenamefont
  {Yang}}]{zheng2012delocalization}%
  \BibitemOpen
  \bibfield  {author} {\bibinfo {author} {\bibfnamefont {X.}~\bibnamefont
  {Zheng}}, \bibinfo {author} {\bibfnamefont {M.}~\bibnamefont {Liu}}, \bibinfo
  {author} {\bibfnamefont {E.~R.}\ \bibnamefont {Johnson}}, \bibinfo {author}
  {\bibfnamefont {J.}~\bibnamefont {Contreras-Garc{\'\i}a}}, \ and\ \bibinfo
  {author} {\bibfnamefont {W.}~\bibnamefont {Yang}},\ }\bibfield  {title}
  {\enquote {\bibinfo {title} {Delocalization error of density-functional
  approximations: A distinct manifestation in hydrogen molecular chains},}\
  }\href@noop {} {\bibfield  {journal} {\bibinfo  {journal} {J. Chem. Phys.}\
  }\textbf {\bibinfo {volume} {137}},\ \bibinfo {pages} {214106} (\bibinfo
  {year} {2012})}\BibitemShut {NoStop}%
\bibitem [{\citenamefont {Dwyer}\ and\ \citenamefont
  {Tozer}(2011)}]{dwyer2011dispersion}%
  \BibitemOpen
  \bibfield  {author} {\bibinfo {author} {\bibfnamefont {A.~D.}\ \bibnamefont
  {Dwyer}}\ and\ \bibinfo {author} {\bibfnamefont {D.~J.}\ \bibnamefont
  {Tozer}},\ }\bibfield  {title} {\enquote {\bibinfo {title} {Dispersion,
  static correlation, and delocalisation errors in density functional theory:
  An electrostatic theorem perspective},}\ }\href@noop {} {\bibfield  {journal}
  {\bibinfo  {journal} {J. Chem. Phys.}\ }\textbf {\bibinfo {volume} {135}}
  (\bibinfo {year} {2011})}\BibitemShut {NoStop}%
\bibitem [{\citenamefont {Li}\ \emph {et~al.}(2018)\citenamefont {Li},
  \citenamefont {Zheng}, \citenamefont {Su},\ and\ \citenamefont
  {Yang}}]{li2018localized}%
  \BibitemOpen
  \bibfield  {author} {\bibinfo {author} {\bibfnamefont {C.}~\bibnamefont
  {Li}}, \bibinfo {author} {\bibfnamefont {X.}~\bibnamefont {Zheng}}, \bibinfo
  {author} {\bibfnamefont {N.~Q.}\ \bibnamefont {Su}}, \ and\ \bibinfo {author}
  {\bibfnamefont {W.}~\bibnamefont {Yang}},\ }\bibfield  {title} {\enquote
  {\bibinfo {title} {Localized orbital scaling correction for systematic
  elimination of delocalization error in density functional approximations},}\
  }\href@noop {} {\bibfield  {journal} {\bibinfo  {journal} {Natl. Sci. Rev.}\
  }\textbf {\bibinfo {volume} {5}},\ \bibinfo {pages} {203--215} (\bibinfo
  {year} {2018})}\BibitemShut {NoStop}%
\bibitem [{\citenamefont {Johnson}, \citenamefont {Otero-De-La-Roza},\ and\
  \citenamefont {Dale}(2013)}]{johnson2013extreme}%
  \BibitemOpen
  \bibfield  {author} {\bibinfo {author} {\bibfnamefont {E.~R.}\ \bibnamefont
  {Johnson}}, \bibinfo {author} {\bibfnamefont {A.}~\bibnamefont
  {Otero-De-La-Roza}}, \ and\ \bibinfo {author} {\bibfnamefont {S.~G.}\
  \bibnamefont {Dale}},\ }\bibfield  {title} {\enquote {\bibinfo {title}
  {Extreme density-driven delocalization error for a model solvated-electron
  system},}\ }\href@noop {} {\bibfield  {journal} {\bibinfo  {journal} {J.
  Chem. Phys.}\ }\textbf {\bibinfo {volume} {139}},\ \bibinfo {pages} {184116}
  (\bibinfo {year} {2013})}\BibitemShut {NoStop}%
\bibitem [{\citenamefont {Ruzsinszky}\ \emph {et~al.}(2006)\citenamefont
  {Ruzsinszky}, \citenamefont {Perdew}, \citenamefont {Csonka}, \citenamefont
  {Vydrov},\ and\ \citenamefont {Scuseria}}]{ruzsinszky2006spurious}%
  \BibitemOpen
  \bibfield  {author} {\bibinfo {author} {\bibfnamefont {A.}~\bibnamefont
  {Ruzsinszky}}, \bibinfo {author} {\bibfnamefont {J.~P.}\ \bibnamefont
  {Perdew}}, \bibinfo {author} {\bibfnamefont {G.~I.}\ \bibnamefont {Csonka}},
  \bibinfo {author} {\bibfnamefont {O.~A.}\ \bibnamefont {Vydrov}}, \ and\
  \bibinfo {author} {\bibfnamefont {G.~E.}\ \bibnamefont {Scuseria}},\
  }\bibfield  {title} {\enquote {\bibinfo {title} {Spurious fractional charge
  on dissociated atoms: Pervasive and resilient self-interaction error of
  common density functionals},}\ }\href {\doibase 10.1063/1.2387954} {\bibfield
   {journal} {\bibinfo  {journal} {J. Chem. Phys.}\ }\textbf {\bibinfo {volume}
  {125}},\ \bibinfo {eid} {194112} (\bibinfo {year} {2006})}\BibitemShut
  {NoStop}%
\bibitem [{\citenamefont {Becke}\ and\ \citenamefont
  {Roussel}(1989)}]{becke1989exchange}%
  \BibitemOpen
  \bibfield  {author} {\bibinfo {author} {\bibfnamefont {A.~D.}\ \bibnamefont
  {Becke}}\ and\ \bibinfo {author} {\bibfnamefont {M.~R.}\ \bibnamefont
  {Roussel}},\ }\bibfield  {title} {\enquote {\bibinfo {title} {Exchange holes
  in inhomogeneous systems: A coordinate-space model},}\ }\href {\doibase
  10.1103/PhysRevA.39.3761} {\bibfield  {journal} {\bibinfo  {journal} {Phys.
  Rev. A}\ }\textbf {\bibinfo {volume} {39}},\ \bibinfo {pages} {3761}
  (\bibinfo {year} {1989})}\BibitemShut {NoStop}%
\bibitem [{\citenamefont {Becke}(1983)}]{becke1983hartree}%
  \BibitemOpen
  \bibfield  {author} {\bibinfo {author} {\bibfnamefont {A.}~\bibnamefont
  {Becke}},\ }\bibfield  {title} {\enquote {\bibinfo {title} {{H}artree--{F}ock
  exchange energy of an inhomogeneous electron gas},}\ }\href {\doibase
  10.1002/qua.560230605} {\bibfield  {journal} {\bibinfo  {journal} {Int. J.
  Quantum Chem.}\ }\textbf {\bibinfo {volume} {23}},\ \bibinfo {pages}
  {1915--1922} (\bibinfo {year} {1983})}\BibitemShut {NoStop}%
\bibitem [{\citenamefont {Tsuneda}, \citenamefont {Kamiya},\ and\ \citenamefont
  {Hirao}(2003)}]{tsuneda2003regional}%
  \BibitemOpen
  \bibfield  {author} {\bibinfo {author} {\bibfnamefont {T.}~\bibnamefont
  {Tsuneda}}, \bibinfo {author} {\bibfnamefont {M.}~\bibnamefont {Kamiya}}, \
  and\ \bibinfo {author} {\bibfnamefont {K.}~\bibnamefont {Hirao}},\ }\bibfield
   {title} {\enquote {\bibinfo {title} {Regional self-interaction correction of
  density functional theory},}\ }\href@noop {} {\bibfield  {journal} {\bibinfo
  {journal} {J. Comput. Chem.}\ }\textbf {\bibinfo {volume} {24}},\ \bibinfo
  {pages} {1592--1598} (\bibinfo {year} {2003})}\BibitemShut {NoStop}%
\bibitem [{\citenamefont {Tsuneda}\ and\ \citenamefont
  {Hirao}(2014)}]{tsuneda2014self}%
  \BibitemOpen
  \bibfield  {author} {\bibinfo {author} {\bibfnamefont {T.}~\bibnamefont
  {Tsuneda}}\ and\ \bibinfo {author} {\bibfnamefont {K.}~\bibnamefont
  {Hirao}},\ }\bibfield  {title} {\enquote {\bibinfo {title} {Self-interaction
  corrections in density functional theory},}\ }\href@noop {} {\bibfield
  {journal} {\bibinfo  {journal} {J. Chem. Phys.}\ }\textbf {\bibinfo {volume}
  {140}},\ \bibinfo {pages} {18A513} (\bibinfo {year} {2014})}\BibitemShut
  {NoStop}%
\bibitem [{\citenamefont {Jaramillo}, \citenamefont {Scuseria},\ and\
  \citenamefont {Ernzerhof}(2003)}]{jaramillo2003local}%
  \BibitemOpen
  \bibfield  {author} {\bibinfo {author} {\bibfnamefont {J.}~\bibnamefont
  {Jaramillo}}, \bibinfo {author} {\bibfnamefont {G.~E.}\ \bibnamefont
  {Scuseria}}, \ and\ \bibinfo {author} {\bibfnamefont {M.}~\bibnamefont
  {Ernzerhof}},\ }\bibfield  {title} {\enquote {\bibinfo {title} {Local hybrid
  functionals},}\ }\href@noop {} {\bibfield  {journal} {\bibinfo  {journal} {J.
  Chem. Phys.}\ }\textbf {\bibinfo {volume} {118}},\ \bibinfo {pages}
  {1068--1073} (\bibinfo {year} {2003})}\BibitemShut {NoStop}%
\bibitem [{\citenamefont {Kaupp}, \citenamefont {Bahmann},\ and\ \citenamefont
  {Arbuznikov}(2007)}]{kaupp2007local}%
  \BibitemOpen
  \bibfield  {author} {\bibinfo {author} {\bibfnamefont {M.}~\bibnamefont
  {Kaupp}}, \bibinfo {author} {\bibfnamefont {H.}~\bibnamefont {Bahmann}}, \
  and\ \bibinfo {author} {\bibfnamefont {A.~V.}\ \bibnamefont {Arbuznikov}},\
  }\bibfield  {title} {\enquote {\bibinfo {title} {Local hybrid functionals: An
  assessment for thermochemical kinetics},}\ }\href {\doibase
  10.1063/1.2795700} {\bibfield  {journal} {\bibinfo  {journal} {J. Chem.
  Phys.}\ }\textbf {\bibinfo {volume} {127}},\ \bibinfo {pages} {194102}
  (\bibinfo {year} {2007})}\BibitemShut {NoStop}%
\bibitem [{\citenamefont {Schmidt}\ \emph {et~al.}(2014)\citenamefont
  {Schmidt}, \citenamefont {Kraisler}, \citenamefont {Makmal}, \citenamefont
  {Kronik},\ and\ \citenamefont {K{\"u}mmel}}]{schmidt2014self}%
  \BibitemOpen
  \bibfield  {author} {\bibinfo {author} {\bibfnamefont {T.}~\bibnamefont
  {Schmidt}}, \bibinfo {author} {\bibfnamefont {E.}~\bibnamefont {Kraisler}},
  \bibinfo {author} {\bibfnamefont {A.}~\bibnamefont {Makmal}}, \bibinfo
  {author} {\bibfnamefont {L.}~\bibnamefont {Kronik}}, \ and\ \bibinfo {author}
  {\bibfnamefont {S.}~\bibnamefont {K{\"u}mmel}},\ }\bibfield  {title}
  {\enquote {\bibinfo {title} {A self-interaction-free local hybrid functional:
  Accurate binding energies vis-{\`a}-vis accurate ionization potentials from
  {K}ohn-{S}ham eigenvalues},}\ }\href@noop {} {\bibfield  {journal} {\bibinfo
  {journal} {J. Chem. Phys.}\ }\textbf {\bibinfo {volume} {140}},\ \bibinfo
  {pages} {18A510} (\bibinfo {year} {2014})}\BibitemShut {NoStop}%
\bibitem [{\citenamefont {Latter}(1955)}]{latter1955atomic}%
  \BibitemOpen
  \bibfield  {author} {\bibinfo {author} {\bibfnamefont {R.}~\bibnamefont
  {Latter}},\ }\bibfield  {title} {\enquote {\bibinfo {title} {Atomic energy
  levels for the {Thomas-Fermi} and {Thomas-Fermi-Dirac} potential},}\ }\href
  {\doibase 10.1103/PhysRev.99.510} {\bibfield  {journal} {\bibinfo  {journal}
  {Phys. Rev.}\ }\textbf {\bibinfo {volume} {99}},\ \bibinfo {pages} {510}
  (\bibinfo {year} {1955})}\BibitemShut {NoStop}%
\bibitem [{\citenamefont {Dabo}, \citenamefont {Ferretti},\ and\ \citenamefont
  {Marzari}(2014)}]{dabo2014piecewise}%
  \BibitemOpen
  \bibfield  {author} {\bibinfo {author} {\bibfnamefont {I.}~\bibnamefont
  {Dabo}}, \bibinfo {author} {\bibfnamefont {A.}~\bibnamefont {Ferretti}}, \
  and\ \bibinfo {author} {\bibfnamefont {N.}~\bibnamefont {Marzari}},\
  }\enquote {\bibinfo {title} {Piecewise linearity and spectroscopic properties
  from {Koopmans}-compliant functionals},}\ in\ \href {\doibase
  10.1007/128_2013_504} {\emph {\bibinfo {booktitle} {First Principles
  Approaches to Spectroscopic Properties of Complex Materials}}},\ \bibinfo
  {editor} {edited by\ \bibinfo {editor} {\bibfnamefont {C.}~\bibnamefont {{Di
  Valentin}}}, \bibinfo {editor} {\bibfnamefont {S.}~\bibnamefont {Botti}}, \
  and\ \bibinfo {editor} {\bibfnamefont {M.}~\bibnamefont {Cococcioni}}}\
  (\bibinfo  {publisher} {Springer Berlin Heidelberg},\ \bibinfo {address}
  {Berlin, Heidelberg},\ \bibinfo {year} {2014})\ pp.\ \bibinfo {pages}
  {193--233}\BibitemShut {NoStop}%
\bibitem [{\citenamefont {Borghi}\ \emph {et~al.}(2014)\citenamefont {Borghi},
  \citenamefont {Ferretti}, \citenamefont {Nguyen}, \citenamefont {Dabo},\ and\
  \citenamefont {Marzari}}]{borghi2014koopmans}%
  \BibitemOpen
  \bibfield  {author} {\bibinfo {author} {\bibfnamefont {G.}~\bibnamefont
  {Borghi}}, \bibinfo {author} {\bibfnamefont {A.}~\bibnamefont {Ferretti}},
  \bibinfo {author} {\bibfnamefont {N.~L.}\ \bibnamefont {Nguyen}}, \bibinfo
  {author} {\bibfnamefont {I.}~\bibnamefont {Dabo}}, \ and\ \bibinfo {author}
  {\bibfnamefont {N.}~\bibnamefont {Marzari}},\ }\bibfield  {title} {\enquote
  {\bibinfo {title} {{K}oopmans-compliant functionals and their performance
  against reference molecular data},}\ }\href {\doibase
  10.1103/PhysRevB.90.075135} {\bibfield  {journal} {\bibinfo  {journal} {Phys.
  Rev. B}\ }\textbf {\bibinfo {volume} {90}},\ \bibinfo {pages} {075135}
  (\bibinfo {year} {2014})}\BibitemShut {NoStop}%
\bibitem [{\citenamefont {Pemmaraju}\ \emph {et~al.}(2007)\citenamefont
  {Pemmaraju}, \citenamefont {Archer}, \citenamefont {S{\'a}nchez-Portal},\
  and\ \citenamefont {Sanvito}}]{pemmaraju2007atomic}%
  \BibitemOpen
  \bibfield  {author} {\bibinfo {author} {\bibfnamefont {C.~D.}\ \bibnamefont
  {Pemmaraju}}, \bibinfo {author} {\bibfnamefont {T.}~\bibnamefont {Archer}},
  \bibinfo {author} {\bibfnamefont {D.}~\bibnamefont {S{\'a}nchez-Portal}}, \
  and\ \bibinfo {author} {\bibfnamefont {S.}~\bibnamefont {Sanvito}},\
  }\bibfield  {title} {\enquote {\bibinfo {title} {Atomic-orbital-based
  approximate self-interaction correction scheme for molecules and solids},}\
  }\href@noop {} {\bibfield  {journal} {\bibinfo  {journal} {Phys. Rev. B}\
  }\textbf {\bibinfo {volume} {75}},\ \bibinfo {pages} {045101} (\bibinfo
  {year} {2007})}\BibitemShut {NoStop}%
\bibitem [{\citenamefont {{Li Manni}}\ \emph {et~al.}(2014)\citenamefont {{Li
  Manni}}, \citenamefont {Carlson}, \citenamefont {Luo}, \citenamefont {Ma},
  \citenamefont {Olsen}, \citenamefont {Truhlar},\ and\ \citenamefont
  {Gagliardi}}]{li2014multiconfiguration}%
  \BibitemOpen
  \bibfield  {author} {\bibinfo {author} {\bibfnamefont {G.}~\bibnamefont {{Li
  Manni}}}, \bibinfo {author} {\bibfnamefont {R.~K.}\ \bibnamefont {Carlson}},
  \bibinfo {author} {\bibfnamefont {S.}~\bibnamefont {Luo}}, \bibinfo {author}
  {\bibfnamefont {D.}~\bibnamefont {Ma}}, \bibinfo {author} {\bibfnamefont
  {J.}~\bibnamefont {Olsen}}, \bibinfo {author} {\bibfnamefont {D.~G.}\
  \bibnamefont {Truhlar}}, \ and\ \bibinfo {author} {\bibfnamefont
  {L.}~\bibnamefont {Gagliardi}},\ }\bibfield  {title} {\enquote {\bibinfo
  {title} {Multiconfiguration pair-density functional theory},}\ }\href@noop {}
  {\bibfield  {journal} {\bibinfo  {journal} {J. Chem. Theory Comput.}\
  }\textbf {\bibinfo {volume} {10}},\ \bibinfo {pages} {3669--3680} (\bibinfo
  {year} {2014})}\BibitemShut {NoStop}%
\bibitem [{\citenamefont {Su}, \citenamefont {Mahler},\ and\ \citenamefont
  {Yang}(2020)}]{su2020preserving}%
  \BibitemOpen
  \bibfield  {author} {\bibinfo {author} {\bibfnamefont {N.~Q.}\ \bibnamefont
  {Su}}, \bibinfo {author} {\bibfnamefont {A.}~\bibnamefont {Mahler}}, \ and\
  \bibinfo {author} {\bibfnamefont {W.}~\bibnamefont {Yang}},\ }\bibfield
  {title} {\enquote {\bibinfo {title} {Preserving symmetry and degeneracy in
  the localized orbital scaling correction approach},}\ }\href@noop {}
  {\bibfield  {journal} {\bibinfo  {journal} {J. Phys. Chem. Lett.}\ }\textbf
  {\bibinfo {volume} {11}},\ \bibinfo {pages} {1528--1535} (\bibinfo {year}
  {2020})}\BibitemShut {NoStop}%
\bibitem [{\citenamefont {Janesko}(2021)}]{janesko2021replacing}%
  \BibitemOpen
  \bibfield  {author} {\bibinfo {author} {\bibfnamefont {B.~G.}\ \bibnamefont
  {Janesko}},\ }\bibfield  {title} {\enquote {\bibinfo {title} {Replacing
  hybrid density functional theory: motivation and recent advances},}\
  }\href@noop {} {\bibfield  {journal} {\bibinfo  {journal} {Chem. Soc. Rev.}\
  } (\bibinfo {year} {2021})}\BibitemShut {NoStop}%
\bibitem [{\citenamefont {Nagai}, \citenamefont {Akashi},\ and\ \citenamefont
  {Sugino}(2020)}]{nagai2020completing}%
  \BibitemOpen
  \bibfield  {author} {\bibinfo {author} {\bibfnamefont {R.}~\bibnamefont
  {Nagai}}, \bibinfo {author} {\bibfnamefont {R.}~\bibnamefont {Akashi}}, \
  and\ \bibinfo {author} {\bibfnamefont {O.}~\bibnamefont {Sugino}},\
  }\bibfield  {title} {\enquote {\bibinfo {title} {Completing density
  functional theory by machine learning hidden messages from molecules},}\
  }\href@noop {} {\bibfield  {journal} {\bibinfo  {journal} {NPJ Comput.
  Mater.}\ }\textbf {\bibinfo {volume} {6}},\ \bibinfo {pages} {43} (\bibinfo
  {year} {2020})}\BibitemShut {NoStop}%
\bibitem [{\citenamefont {Kirkpatrick}\ \emph {et~al.}(2021)\citenamefont
  {Kirkpatrick}, \citenamefont {McMorrow}, \citenamefont {Turban},
  \citenamefont {Gaunt}, \citenamefont {Spencer}, \citenamefont {Matthews},
  \citenamefont {Obika}, \citenamefont {Thiry}, \citenamefont {Fortunato},
  \citenamefont {Pfau} \emph {et~al.}}]{kirkpatrick2021pushing}%
  \BibitemOpen
  \bibfield  {author} {\bibinfo {author} {\bibfnamefont {J.}~\bibnamefont
  {Kirkpatrick}}, \bibinfo {author} {\bibfnamefont {B.}~\bibnamefont
  {McMorrow}}, \bibinfo {author} {\bibfnamefont {D.~H.}\ \bibnamefont
  {Turban}}, \bibinfo {author} {\bibfnamefont {A.~L.}\ \bibnamefont {Gaunt}},
  \bibinfo {author} {\bibfnamefont {J.~S.}\ \bibnamefont {Spencer}}, \bibinfo
  {author} {\bibfnamefont {A.~G.}\ \bibnamefont {Matthews}}, \bibinfo {author}
  {\bibfnamefont {A.}~\bibnamefont {Obika}}, \bibinfo {author} {\bibfnamefont
  {L.}~\bibnamefont {Thiry}}, \bibinfo {author} {\bibfnamefont
  {M.}~\bibnamefont {Fortunato}}, \bibinfo {author} {\bibfnamefont
  {D.}~\bibnamefont {Pfau}},  \emph {et~al.},\ }\bibfield  {title} {\enquote
  {\bibinfo {title} {Pushing the frontiers of density functionals by solving
  the fractional electron problem},}\ }\href@noop {} {\bibfield  {journal}
  {\bibinfo  {journal} {Science}\ }\textbf {\bibinfo {volume} {374}},\ \bibinfo
  {pages} {1385--1389} (\bibinfo {year} {2021})}\BibitemShut {NoStop}%
\bibitem [{\citenamefont {Bogojeski}\ \emph {et~al.}(2020)\citenamefont
  {Bogojeski}, \citenamefont {Vogt-Maranto}, \citenamefont {Tuckerman},
  \citenamefont {M{\"u}ller},\ and\ \citenamefont
  {Burke}}]{bogojeski2020quantum}%
  \BibitemOpen
  \bibfield  {author} {\bibinfo {author} {\bibfnamefont {M.}~\bibnamefont
  {Bogojeski}}, \bibinfo {author} {\bibfnamefont {L.}~\bibnamefont
  {Vogt-Maranto}}, \bibinfo {author} {\bibfnamefont {M.~E.}\ \bibnamefont
  {Tuckerman}}, \bibinfo {author} {\bibfnamefont {K.-R.}\ \bibnamefont
  {M{\"u}ller}}, \ and\ \bibinfo {author} {\bibfnamefont {K.}~\bibnamefont
  {Burke}},\ }\bibfield  {title} {\enquote {\bibinfo {title} {Quantum chemical
  accuracy from density functional approximations via machine learning},}\
  }\href@noop {} {\bibfield  {journal} {\bibinfo  {journal} {Nat. Commun.}\
  }\textbf {\bibinfo {volume} {11}},\ \bibinfo {pages} {5223} (\bibinfo {year}
  {2020})}\BibitemShut {NoStop}%
\bibitem [{\citenamefont {Dick}\ and\ \citenamefont
  {Fernandez-Serra}(2021)}]{dick2021highly}%
  \BibitemOpen
  \bibfield  {author} {\bibinfo {author} {\bibfnamefont {S.}~\bibnamefont
  {Dick}}\ and\ \bibinfo {author} {\bibfnamefont {M.}~\bibnamefont
  {Fernandez-Serra}},\ }\bibfield  {title} {\enquote {\bibinfo {title} {Highly
  accurate and constrained density functional obtained with differentiable
  programming},}\ }\href@noop {} {\bibfield  {journal} {\bibinfo  {journal}
  {Phys. Rev. B}\ }\textbf {\bibinfo {volume} {104}},\ \bibinfo {pages}
  {L161109} (\bibinfo {year} {2021})}\BibitemShut {NoStop}%
\bibitem [{\citenamefont {Zope}\ \emph {et~al.}(2019)\citenamefont {Zope},
  \citenamefont {Yamamoto}, \citenamefont {Diaz}, \citenamefont {Baruah},
  \citenamefont {Peralta}, \citenamefont {Jackson}, \citenamefont {Santra},\
  and\ \citenamefont {Perdew}}]{zope2019step}%
  \BibitemOpen
  \bibfield  {author} {\bibinfo {author} {\bibfnamefont {R.~R.}\ \bibnamefont
  {Zope}}, \bibinfo {author} {\bibfnamefont {Y.}~\bibnamefont {Yamamoto}},
  \bibinfo {author} {\bibfnamefont {C.~M.}\ \bibnamefont {Diaz}}, \bibinfo
  {author} {\bibfnamefont {T.}~\bibnamefont {Baruah}}, \bibinfo {author}
  {\bibfnamefont {J.~E.}\ \bibnamefont {Peralta}}, \bibinfo {author}
  {\bibfnamefont {K.~A.}\ \bibnamefont {Jackson}}, \bibinfo {author}
  {\bibfnamefont {B.}~\bibnamefont {Santra}}, \ and\ \bibinfo {author}
  {\bibfnamefont {J.~P.}\ \bibnamefont {Perdew}},\ }\bibfield  {title}
  {\enquote {\bibinfo {title} {A step in the direction of resolving the paradox
  of {P}erdew-{Z}unger self-interaction correction},}\ }\href@noop {}
  {\bibfield  {journal} {\bibinfo  {journal} {J. Chem. Phys.}\ }\textbf
  {\bibinfo {volume} {151}},\ \bibinfo {pages} {214108} (\bibinfo {year}
  {2019})}\BibitemShut {NoStop}%
\bibitem [{\citenamefont {Yamamoto}\ \emph {et~al.}(2023)\citenamefont
  {Yamamoto}, \citenamefont {Baruah}, \citenamefont {Chang}, \citenamefont
  {Romero},\ and\ \citenamefont {Zope}}]{yamamoto2023self}%
  \BibitemOpen
  \bibfield  {author} {\bibinfo {author} {\bibfnamefont {Y.}~\bibnamefont
  {Yamamoto}}, \bibinfo {author} {\bibfnamefont {T.}~\bibnamefont {Baruah}},
  \bibinfo {author} {\bibfnamefont {P.-H.}\ \bibnamefont {Chang}}, \bibinfo
  {author} {\bibfnamefont {S.}~\bibnamefont {Romero}}, \ and\ \bibinfo {author}
  {\bibfnamefont {R.~R.}\ \bibnamefont {Zope}},\ }\bibfield  {title} {\enquote
  {\bibinfo {title} {Self-consistent implementation of locally scaled
  self-interaction-correction method},}\ }\href@noop {} {\bibfield  {journal}
  {\bibinfo  {journal} {J. Chem. Phys.}\ }\textbf {\bibinfo {volume} {158}}
  (\bibinfo {year} {2023})}\BibitemShut {NoStop}%
\bibitem [{\citenamefont {Akter}\ \emph {et~al.}(2020)\citenamefont {Akter},
  \citenamefont {Yamamoto}, \citenamefont {Diaz}, \citenamefont {Jackson},
  \citenamefont {Zope},\ and\ \citenamefont {Baruah}}]{waterpolarizability}%
  \BibitemOpen
  \bibfield  {author} {\bibinfo {author} {\bibfnamefont {S.}~\bibnamefont
  {Akter}}, \bibinfo {author} {\bibfnamefont {Y.}~\bibnamefont {Yamamoto}},
  \bibinfo {author} {\bibfnamefont {C.~M.}\ \bibnamefont {Diaz}}, \bibinfo
  {author} {\bibfnamefont {K.~A.}\ \bibnamefont {Jackson}}, \bibinfo {author}
  {\bibfnamefont {R.~R.}\ \bibnamefont {Zope}}, \ and\ \bibinfo {author}
  {\bibfnamefont {T.}~\bibnamefont {Baruah}},\ }\bibfield  {title} {\enquote
  {\bibinfo {title} {Study of self-interaction errors in density functional
  predictions of dipole polarizabilities and ionization energies of water
  clusters using {P}erdew-{Z}unger and locally scaled self-interaction
  corrected methods},}\ }\href@noop {} {\bibfield  {journal} {\bibinfo
  {journal} {J. Chem. Phys.}\ }\textbf {\bibinfo {volume} {153}},\ \bibinfo
  {pages} {164304} (\bibinfo {year} {2020})}\BibitemShut {NoStop}%
\bibitem [{\citenamefont {Akter}\ \emph
  {et~al.}(2021{\natexlab{a}})\citenamefont {Akter}, \citenamefont {Yamamoto},
  \citenamefont {Zope},\ and\ \citenamefont {Baruah}}]{akter2021static}%
  \BibitemOpen
  \bibfield  {author} {\bibinfo {author} {\bibfnamefont {S.}~\bibnamefont
  {Akter}}, \bibinfo {author} {\bibfnamefont {Y.}~\bibnamefont {Yamamoto}},
  \bibinfo {author} {\bibfnamefont {R.~R.}\ \bibnamefont {Zope}}, \ and\
  \bibinfo {author} {\bibfnamefont {T.}~\bibnamefont {Baruah}},\ }\bibfield
  {title} {\enquote {\bibinfo {title} {Static dipole polarizabilities of
  polyacenes using self-interaction-corrected density functional
  approximations},}\ }\href@noop {} {\bibfield  {journal} {\bibinfo  {journal}
  {J. Chem. Phys.}\ }\textbf {\bibinfo {volume} {154}},\ \bibinfo {pages}
  {114305} (\bibinfo {year} {2021}{\natexlab{a}})}\BibitemShut {NoStop}%
\bibitem [{\citenamefont {Mishra}\ \emph
  {et~al.}(2022{\natexlab{a}})\citenamefont {Mishra}, \citenamefont {Yamamoto},
  \citenamefont {Chang}, \citenamefont {Nguyen}, \citenamefont {Peralta},
  \citenamefont {Baruah},\ and\ \citenamefont
  {Zope}}]{doi:10.1021/acs.jpca.1c10354}%
  \BibitemOpen
  \bibfield  {author} {\bibinfo {author} {\bibfnamefont {P.}~\bibnamefont
  {Mishra}}, \bibinfo {author} {\bibfnamefont {Y.}~\bibnamefont {Yamamoto}},
  \bibinfo {author} {\bibfnamefont {P.-H.}\ \bibnamefont {Chang}}, \bibinfo
  {author} {\bibfnamefont {D.~B.}\ \bibnamefont {Nguyen}}, \bibinfo {author}
  {\bibfnamefont {J.~E.}\ \bibnamefont {Peralta}}, \bibinfo {author}
  {\bibfnamefont {T.}~\bibnamefont {Baruah}}, \ and\ \bibinfo {author}
  {\bibfnamefont {R.~R.}\ \bibnamefont {Zope}},\ }\bibfield  {title} {\enquote
  {\bibinfo {title} {Study of self-interaction errors in density functional
  calculations of magnetic exchange coupling constants using three
  self-interaction correction methods},}\ }\href {\doibase
  10.1021/acs.jpca.1c10354} {\bibfield  {journal} {\bibinfo  {journal} {J.
  Phys. Chem. A}\ }\textbf {\bibinfo {volume} {126}},\ \bibinfo {pages}
  {1923--1935} (\bibinfo {year} {2022}{\natexlab{a}})}\BibitemShut {NoStop}%
\bibitem [{\citenamefont {Mishra}\ \emph
  {et~al.}(2022{\natexlab{b}})\citenamefont {Mishra}, \citenamefont {Yamamoto},
  \citenamefont {Johnson}, \citenamefont {Jackson}, \citenamefont {Zope},\ and\
  \citenamefont {Baruah}}]{mishra2022study}%
  \BibitemOpen
  \bibfield  {author} {\bibinfo {author} {\bibfnamefont {P.}~\bibnamefont
  {Mishra}}, \bibinfo {author} {\bibfnamefont {Y.}~\bibnamefont {Yamamoto}},
  \bibinfo {author} {\bibfnamefont {J.~K.}\ \bibnamefont {Johnson}}, \bibinfo
  {author} {\bibfnamefont {K.~A.}\ \bibnamefont {Jackson}}, \bibinfo {author}
  {\bibfnamefont {R.~R.}\ \bibnamefont {Zope}}, \ and\ \bibinfo {author}
  {\bibfnamefont {T.}~\bibnamefont {Baruah}},\ }\bibfield  {title} {\enquote
  {\bibinfo {title} {Study of self-interaction-errors in barrier heights using
  locally scaled and {P}erdew--{Z}unger self-interaction methods},}\ }\href
  {\doibase 10.1063/5.0070893} {\bibfield  {journal} {\bibinfo  {journal} {J.
  Chem. Phys.}\ }\textbf {\bibinfo {volume} {156}},\ \bibinfo {pages} {014306}
  (\bibinfo {year} {2022}{\natexlab{b}})}\BibitemShut {NoStop}%
\bibitem [{\citenamefont {Akter}\ \emph
  {et~al.}(2021{\natexlab{b}})\citenamefont {Akter}, \citenamefont {Tellez},
  \citenamefont {Sharkas}, \citenamefont {Peralta}, \citenamefont {Jackson},
  \citenamefont {Baruah},\ and\ \citenamefont {Zope}}]{akter2021well}%
  \BibitemOpen
  \bibfield  {author} {\bibinfo {author} {\bibfnamefont {S.}~\bibnamefont
  {Akter}}, \bibinfo {author} {\bibfnamefont {J.~A.~V.}\ \bibnamefont
  {Tellez}}, \bibinfo {author} {\bibfnamefont {K.}~\bibnamefont {Sharkas}},
  \bibinfo {author} {\bibfnamefont {J.}~\bibnamefont {Peralta}}, \bibinfo
  {author} {\bibfnamefont {K.}~\bibnamefont {Jackson}}, \bibinfo {author}
  {\bibfnamefont {T.}~\bibnamefont {Baruah}}, \ and\ \bibinfo {author}
  {\bibfnamefont {R.}~\bibnamefont {Zope}},\ }\bibfield  {title} {\enquote
  {\bibinfo {title} {How well do self-interaction corrections repair the
  over-estimation of molecular polarizabilities in density functional
  calculations?}}\ }\href@noop {} {\bibfield  {journal} {\bibinfo  {journal}
  {Phys. Chem. Chem. Phys.}\ } (\bibinfo {year}
  {2021}{\natexlab{b}})}\BibitemShut {NoStop}%
\bibitem [{\citenamefont {Romero}, \citenamefont {Baruah},\ and\ \citenamefont
  {Zope}(2023)}]{romero2023spin}%
  \BibitemOpen
  \bibfield  {author} {\bibinfo {author} {\bibfnamefont {S.}~\bibnamefont
  {Romero}}, \bibinfo {author} {\bibfnamefont {T.}~\bibnamefont {Baruah}}, \
  and\ \bibinfo {author} {\bibfnamefont {R.~R.}\ \bibnamefont {Zope}},\
  }\bibfield  {title} {\enquote {\bibinfo {title} {Spin-state gaps and
  self-interaction-corrected density functional approximations: Octahedral {Fe
  (II)} complexes as case study},}\ }\href@noop {} {\bibfield  {journal}
  {\bibinfo  {journal} {J. Chem. Phys.}\ }\textbf {\bibinfo {volume} {158}}
  (\bibinfo {year} {2023})}\BibitemShut {NoStop}%
\bibitem [{\citenamefont {Pederson}, \citenamefont {Ruzsinszky},\ and\
  \citenamefont {Perdew}(2014)}]{pederson2014communication}%
  \BibitemOpen
  \bibfield  {author} {\bibinfo {author} {\bibfnamefont {M.~R.}\ \bibnamefont
  {Pederson}}, \bibinfo {author} {\bibfnamefont {A.}~\bibnamefont
  {Ruzsinszky}}, \ and\ \bibinfo {author} {\bibfnamefont {J.~P.}\ \bibnamefont
  {Perdew}},\ }\bibfield  {title} {\enquote {\bibinfo {title} {Communication:
  Self-interaction correction with unitary invariance in density functional
  theory},}\ }\href {\doibase 10.1063/1.4869581} {\bibfield  {journal}
  {\bibinfo  {journal} {J. Chem. Phys.}\ }\textbf {\bibinfo {volume} {140}},\
  \bibinfo {eid} {121103} (\bibinfo {year} {2014})}\BibitemShut {NoStop}%
\bibitem [{\citenamefont {Luken}\ and\ \citenamefont
  {Beratan}(1982)}]{luken1982localized}%
  \BibitemOpen
  \bibfield  {author} {\bibinfo {author} {\bibfnamefont {W.~L.}\ \bibnamefont
  {Luken}}\ and\ \bibinfo {author} {\bibfnamefont {D.~N.}\ \bibnamefont
  {Beratan}},\ }\bibfield  {title} {\enquote {\bibinfo {title} {Localized
  orbitals and the {F}ermi hole},}\ }\href {\doibase 10.1007/BF00550971}
  {\bibfield  {journal} {\bibinfo  {journal} {Theor. Chem. Acc.}\ }\textbf
  {\bibinfo {volume} {61}},\ \bibinfo {pages} {265--281} (\bibinfo {year}
  {1982})}\BibitemShut {NoStop}%
\bibitem [{\citenamefont {Luken}\ and\ \citenamefont
  {Culberson}(1984)}]{luken1984localized}%
  \BibitemOpen
  \bibfield  {author} {\bibinfo {author} {\bibfnamefont {W.~L.}\ \bibnamefont
  {Luken}}\ and\ \bibinfo {author} {\bibfnamefont {J.~C.}\ \bibnamefont
  {Culberson}},\ }\bibfield  {title} {\enquote {\bibinfo {title} {Localized
  orbitals based on the {F}ermi hole},}\ }\href@noop {} {\bibfield  {journal}
  {\bibinfo  {journal} {Theor. Chem. Acc.}\ }\textbf {\bibinfo {volume} {66}},\
  \bibinfo {pages} {279--293} (\bibinfo {year} {1984})}\BibitemShut {NoStop}%
\bibitem [{\citenamefont {Diaz}, \citenamefont {Baruah},\ and\ \citenamefont
  {Zope}(2021)}]{PhysRevA.103.042811}%
  \BibitemOpen
  \bibfield  {author} {\bibinfo {author} {\bibfnamefont {C.~M.}\ \bibnamefont
  {Diaz}}, \bibinfo {author} {\bibfnamefont {T.}~\bibnamefont {Baruah}}, \ and\
  \bibinfo {author} {\bibfnamefont {R.~R.}\ \bibnamefont {Zope}},\ }\bibfield
  {title} {\enquote {\bibinfo {title} {{F}ermi-{L}{\"o}wdin-orbital
  self-interaction correction using the optimized-effective-potential method
  within the {K}rieger-{L}i-{I}afrate approximation},}\ }\href@noop {}
  {\bibfield  {journal} {\bibinfo  {journal} {Phys. Rev. A}\ }\textbf {\bibinfo
  {volume} {103}},\ \bibinfo {pages} {042811} (\bibinfo {year}
  {2021})}\BibitemShut {NoStop}%
\bibitem [{\citenamefont {Romero}\ \emph {et~al.}(2021)\citenamefont {Romero},
  \citenamefont {Yamamoto}, \citenamefont {Baruah},\ and\ \citenamefont
  {Zope}}]{romero2021local}%
  \BibitemOpen
  \bibfield  {author} {\bibinfo {author} {\bibfnamefont {S.}~\bibnamefont
  {Romero}}, \bibinfo {author} {\bibfnamefont {Y.}~\bibnamefont {Yamamoto}},
  \bibinfo {author} {\bibfnamefont {T.}~\bibnamefont {Baruah}}, \ and\ \bibinfo
  {author} {\bibfnamefont {R.~R.}\ \bibnamefont {Zope}},\ }\bibfield  {title}
  {\enquote {\bibinfo {title} {Local self-interaction correction method with a
  simple scaling factor},}\ }\href@noop {} {\bibfield  {journal} {\bibinfo
  {journal} {Phys. Chem. Chem. Phys.}\ }\textbf {\bibinfo {volume} {23}},\
  \bibinfo {pages} {2406--2418} (\bibinfo {year} {2021})}\BibitemShut {NoStop}%
\bibitem [{\citenamefont {Bhattarai}\ \emph {et~al.}(2021)\citenamefont
  {Bhattarai}, \citenamefont {Santra}, \citenamefont {Wagle}, \citenamefont
  {Yamamoto}, \citenamefont {Zope}, \citenamefont {Ruzsinszky}, \citenamefont
  {Jackson},\ and\ \citenamefont {Perdew}}]{doi:10.1063/5.0041646}%
  \BibitemOpen
  \bibfield  {author} {\bibinfo {author} {\bibfnamefont {P.}~\bibnamefont
  {Bhattarai}}, \bibinfo {author} {\bibfnamefont {B.}~\bibnamefont {Santra}},
  \bibinfo {author} {\bibfnamefont {K.}~\bibnamefont {Wagle}}, \bibinfo
  {author} {\bibfnamefont {Y.}~\bibnamefont {Yamamoto}}, \bibinfo {author}
  {\bibfnamefont {R.~R.}\ \bibnamefont {Zope}}, \bibinfo {author}
  {\bibfnamefont {A.}~\bibnamefont {Ruzsinszky}}, \bibinfo {author}
  {\bibfnamefont {K.~A.}\ \bibnamefont {Jackson}}, \ and\ \bibinfo {author}
  {\bibfnamefont {J.~P.}\ \bibnamefont {Perdew}},\ }\bibfield  {title}
  {\enquote {\bibinfo {title} {Exploring and enhancing the accuracy of
  interior-scaled {P}erdew-{Z}unger self-interaction correction},}\ }\href@noop
  {} {\bibfield  {journal} {\bibinfo  {journal} {J. Chem. Phys.}\ }\textbf
  {\bibinfo {volume} {154}},\ \bibinfo {pages} {094105} (\bibinfo {year}
  {2021})}\BibitemShut {NoStop}%
\bibitem [{\citenamefont {Leonard}\ and\ \citenamefont
  {Luken}(1982)}]{leonard1982quadratically}%
  \BibitemOpen
  \bibfield  {author} {\bibinfo {author} {\bibfnamefont {J.~M.}\ \bibnamefont
  {Leonard}}\ and\ \bibinfo {author} {\bibfnamefont {W.~L.}\ \bibnamefont
  {Luken}},\ }\bibfield  {title} {\enquote {\bibinfo {title} {Quadratically
  convergent calculation of localized molecular orbitals},}\ }\href@noop {}
  {\bibfield  {journal} {\bibinfo  {journal} {Theor. Chem. Acc.}\ }\textbf
  {\bibinfo {volume} {62}},\ \bibinfo {pages} {107--132} (\bibinfo {year}
  {1982})}\BibitemShut {NoStop}%
\bibitem [{\citenamefont {Lundin}\ and\ \citenamefont
  {Eriksson}(2001)}]{lundin2001novel}%
  \BibitemOpen
  \bibfield  {author} {\bibinfo {author} {\bibfnamefont {U.}~\bibnamefont
  {Lundin}}\ and\ \bibinfo {author} {\bibfnamefont {O.}~\bibnamefont
  {Eriksson}},\ }\bibfield  {title} {\enquote {\bibinfo {title} {Novel method
  of self-interaction corrections in density functional calculations},}\
  }\href@noop {} {\bibfield  {journal} {\bibinfo  {journal} {Int. J. Quantum
  Chem.}\ }\textbf {\bibinfo {volume} {81}},\ \bibinfo {pages} {247--252}
  (\bibinfo {year} {2001})}\BibitemShut {NoStop}%
\bibitem [{\citenamefont {Vydrov}\ \emph {et~al.}(2006)\citenamefont {Vydrov},
  \citenamefont {Scuseria}, \citenamefont {Perdew}, \citenamefont
  {Ruzsinszky},\ and\ \citenamefont {Csonka}}]{vydrov2006scaling}%
  \BibitemOpen
  \bibfield  {author} {\bibinfo {author} {\bibfnamefont {O.~A.}\ \bibnamefont
  {Vydrov}}, \bibinfo {author} {\bibfnamefont {G.~E.}\ \bibnamefont
  {Scuseria}}, \bibinfo {author} {\bibfnamefont {J.~P.}\ \bibnamefont
  {Perdew}}, \bibinfo {author} {\bibfnamefont {A.}~\bibnamefont {Ruzsinszky}},
  \ and\ \bibinfo {author} {\bibfnamefont {G.~I.}\ \bibnamefont {Csonka}},\
  }\bibfield  {title} {\enquote {\bibinfo {title} {Scaling down the
  {P}erdew-{Z}unger self-interaction correction in many-electron regions},}\
  }\href@noop {} {\bibfield  {journal} {\bibinfo  {journal} {J. Chem. Phys.}\
  }\textbf {\bibinfo {volume} {124}},\ \bibinfo {pages} {094108} (\bibinfo
  {year} {2006})}\BibitemShut {NoStop}%
\bibitem [{\citenamefont {Yamamoto}\ \emph {et~al.}(2020)\citenamefont
  {Yamamoto}, \citenamefont {Romero}, \citenamefont {Baruah},\ and\
  \citenamefont {Zope}}]{yamamoto2020improvements}%
  \BibitemOpen
  \bibfield  {author} {\bibinfo {author} {\bibfnamefont {Y.}~\bibnamefont
  {Yamamoto}}, \bibinfo {author} {\bibfnamefont {S.}~\bibnamefont {Romero}},
  \bibinfo {author} {\bibfnamefont {T.}~\bibnamefont {Baruah}}, \ and\ \bibinfo
  {author} {\bibfnamefont {R.~R.}\ \bibnamefont {Zope}},\ }\bibfield  {title}
  {\enquote {\bibinfo {title} {Improvements in the orbitalwise scaling down of
  {P}erdew--{Z}unger self-interaction correction in many-electron regions},}\
  }\href@noop {} {\bibfield  {journal} {\bibinfo  {journal} {J. Chem. Phys.}\
  }\textbf {\bibinfo {volume} {152}},\ \bibinfo {pages} {174112} (\bibinfo
  {year} {2020})}\BibitemShut {NoStop}%
\bibitem [{\citenamefont {Romero}\ \emph {et~al.}(2023)\citenamefont {Romero},
  \citenamefont {Yamamoto}, \citenamefont {Baruah},\ and\ \citenamefont
  {Zope}}]{romero2023complexity}%
  \BibitemOpen
  \bibfield  {author} {\bibinfo {author} {\bibfnamefont {S.}~\bibnamefont
  {Romero}}, \bibinfo {author} {\bibfnamefont {Y.}~\bibnamefont {Yamamoto}},
  \bibinfo {author} {\bibfnamefont {T.}~\bibnamefont {Baruah}}, \ and\ \bibinfo
  {author} {\bibfnamefont {R.~R.}\ \bibnamefont {Zope}},\ }\bibfield  {title}
  {\enquote {\bibinfo {title} {Complexity reduction in self-interaction-free
  density functional calculations using the {F}ermi-{L}{\"o}wdin
  self-interaction correction method},}\ }\href@noop {} {\bibfield  {journal}
  {\bibinfo  {journal} {arXiv preprint arXiv:2308.04664}\ } (\bibinfo {year}
  {2023})}\BibitemShut {NoStop}%
\bibitem [{\citenamefont {L{\"o}wdin}(1950)}]{lowdin1950non}%
  \BibitemOpen
  \bibfield  {author} {\bibinfo {author} {\bibfnamefont {P.-O.}\ \bibnamefont
  {L{\"o}wdin}},\ }\bibfield  {title} {\enquote {\bibinfo {title} {On the
  non-orthogonality problem connected with the use of atomic wave functions in
  the theory of molecules and crystals},}\ }\href@noop {} {\bibfield  {journal}
  {\bibinfo  {journal} {J. Chem. Phys.}\ }\textbf {\bibinfo {volume} {18}},\
  \bibinfo {pages} {365--375} (\bibinfo {year} {1950})}\BibitemShut {NoStop}%
\bibitem [{\citenamefont {hui Yang}, \citenamefont {Pederson},\ and\
  \citenamefont {Perdew}(2017)}]{yang2017full}%
  \BibitemOpen
  \bibfield  {author} {\bibinfo {author} {\bibfnamefont {Z.}~\bibnamefont {hui
  Yang}}, \bibinfo {author} {\bibfnamefont {M.~R.}\ \bibnamefont {Pederson}}, \
  and\ \bibinfo {author} {\bibfnamefont {J.~P.}\ \bibnamefont {Perdew}},\
  }\bibfield  {title} {\enquote {\bibinfo {title} {Full self-consistency in the
  {F}ermi-orbital self-interaction correction},}\ }\href {\doibase
  10.1103/PhysRevA.95.052505} {\bibfield  {journal} {\bibinfo  {journal} {Phys.
  Rev. A}\ }\textbf {\bibinfo {volume} {95}},\ \bibinfo {pages} {052505}
  (\bibinfo {year} {2017})}\BibitemShut {NoStop}%
\bibitem [{\citenamefont {Pederson}(2015)}]{pederson2015fermi}%
  \BibitemOpen
  \bibfield  {author} {\bibinfo {author} {\bibfnamefont {M.~R.}\ \bibnamefont
  {Pederson}},\ }\bibfield  {title} {\enquote {\bibinfo {title} {Fermi orbital
  derivatives in self-interaction corrected density functional theory:
  Applications to closed shell atoms},}\ }\href {\doibase 10.1063/1.4907592}
  {\bibfield  {journal} {\bibinfo  {journal} {J. Chem. Phys.}\ }\textbf
  {\bibinfo {volume} {142}},\ \bibinfo {eid} {064112} (\bibinfo {year}
  {2015})},\ \Eprint {http://arxiv.org/abs/https://doi.org/10.1063/1.4907592}
  {https://doi.org/10.1063/1.4907592} \BibitemShut {NoStop}%
\bibitem [{\citenamefont {Pederson}\ and\ \citenamefont
  {Baruah}(2015)}]{pederson2015self}%
  \BibitemOpen
  \bibfield  {author} {\bibinfo {author} {\bibfnamefont {M.~R.}\ \bibnamefont
  {Pederson}}\ and\ \bibinfo {author} {\bibfnamefont {T.}~\bibnamefont
  {Baruah}},\ }\bibfield  {title} {\enquote {\bibinfo {title} {Self-interaction
  corrections within the {F}ermi-orbital-based formalism},}\ }in\ \href@noop {}
  {\emph {\bibinfo {booktitle} {Advances in Atomic, Molecular, and Optical
  Physics}}},\ Vol.~\bibinfo {volume} {64},\ \bibinfo {editor} {edited by\
  \bibinfo {editor} {\bibfnamefont {E.}~\bibnamefont {Arimondo}}, \bibinfo
  {editor} {\bibfnamefont {C.~C.}\ \bibnamefont {Lin}}, \ and\ \bibinfo
  {editor} {\bibfnamefont {S.~F.}\ \bibnamefont {Yelin}}}\ (\bibinfo
  {publisher} {Elsevier},\ \bibinfo {year} {2015})\ pp.\ \bibinfo {pages}
  {153--180}\BibitemShut {NoStop}%
\bibitem [{\citenamefont {Zope}, \citenamefont {Baruah},\ and\ \citenamefont
  {Jackson}()}]{FLOSICcode}%
  \BibitemOpen
  \bibfield  {author} {\bibinfo {author} {\bibfnamefont {R.~R.}\ \bibnamefont
  {Zope}}, \bibinfo {author} {\bibfnamefont {T.}~\bibnamefont {Baruah}}, \ and\
  \bibinfo {author} {\bibfnamefont {K.~A.}\ \bibnamefont {Jackson}},\ }\href
  {https://flosic.org/} {\enquote {\bibinfo {title} {\uppercase{FLOSIC 0.2}},}\
  }\bibinfo {note} {Based on the NRLMOL code of M. R. Pederson}\BibitemShut
  {NoStop}%
\bibitem [{\citenamefont {Yamamoto}\ \emph {et~al.}()\citenamefont {Yamamoto},
  \citenamefont {Basurto}, \citenamefont {Diaz}, \citenamefont {Zope},\ and\
  \citenamefont {Baruah}}]{FLOSICcodep}%
  \BibitemOpen
  \bibfield  {author} {\bibinfo {author} {\bibfnamefont {Y.}~\bibnamefont
  {Yamamoto}}, \bibinfo {author} {\bibfnamefont {L.}~\bibnamefont {Basurto}},
  \bibinfo {author} {\bibfnamefont {C.~M.}\ \bibnamefont {Diaz}}, \bibinfo
  {author} {\bibfnamefont {R.~R.}\ \bibnamefont {Zope}}, \ and\ \bibinfo
  {author} {\bibfnamefont {T.}~\bibnamefont {Baruah}},\ }\href
  {https://github.com/FLOSIC/PublicRelease_2020/} {\enquote {\bibinfo {title}
  {Flosic software public release},}\ }\bibinfo {note} {Based on the NRLMOL
  code of M. R. Pederson}\BibitemShut {NoStop}%
\bibitem [{\citenamefont {Bhattarai}\ \emph {et~al.}(2020)\citenamefont
  {Bhattarai}, \citenamefont {Wagle}, \citenamefont {Shahi}, \citenamefont
  {Yamamoto}, \citenamefont {Romero}, \citenamefont {Santra}, \citenamefont
  {Zope}, \citenamefont {Peralta}, \citenamefont {Jackson},\ and\ \citenamefont
  {Perdew}}]{bhattarai2020step}%
  \BibitemOpen
  \bibfield  {author} {\bibinfo {author} {\bibfnamefont {P.}~\bibnamefont
  {Bhattarai}}, \bibinfo {author} {\bibfnamefont {K.}~\bibnamefont {Wagle}},
  \bibinfo {author} {\bibfnamefont {C.}~\bibnamefont {Shahi}}, \bibinfo
  {author} {\bibfnamefont {Y.}~\bibnamefont {Yamamoto}}, \bibinfo {author}
  {\bibfnamefont {S.}~\bibnamefont {Romero}}, \bibinfo {author} {\bibfnamefont
  {B.}~\bibnamefont {Santra}}, \bibinfo {author} {\bibfnamefont {R.~R.}\
  \bibnamefont {Zope}}, \bibinfo {author} {\bibfnamefont {J.~E.}\ \bibnamefont
  {Peralta}}, \bibinfo {author} {\bibfnamefont {K.~A.}\ \bibnamefont
  {Jackson}}, \ and\ \bibinfo {author} {\bibfnamefont {J.~P.}\ \bibnamefont
  {Perdew}},\ }\bibfield  {title} {\enquote {\bibinfo {title} {A step in the
  direction of resolving the paradox of {P}erdew--{Z}unger self-interaction
  correction. {II}. gauge consistency of the energy density at three levels of
  approximation},}\ }\href@noop {} {\bibfield  {journal} {\bibinfo  {journal}
  {J. Chem. Phys.}\ }\textbf {\bibinfo {volume} {152}},\ \bibinfo {pages}
  {214109} (\bibinfo {year} {2020})}\BibitemShut {NoStop}%
\bibitem [{\citenamefont {Tao}\ \emph {et~al.}(2008)\citenamefont {Tao},
  \citenamefont {Staroverov}, \citenamefont {Scuseria},\ and\ \citenamefont
  {Perdew}}]{PhysRevA.77.012509}%
  \BibitemOpen
  \bibfield  {author} {\bibinfo {author} {\bibfnamefont {J.}~\bibnamefont
  {Tao}}, \bibinfo {author} {\bibfnamefont {V.~N.}\ \bibnamefont {Staroverov}},
  \bibinfo {author} {\bibfnamefont {G.~E.}\ \bibnamefont {Scuseria}}, \ and\
  \bibinfo {author} {\bibfnamefont {J.~P.}\ \bibnamefont {Perdew}},\ }\bibfield
   {title} {\enquote {\bibinfo {title} {Exact-exchange energy density in the
  gauge of a semilocal density-functional approximation},}\ }\href {\doibase
  10.1103/PhysRevA.77.012509} {\bibfield  {journal} {\bibinfo  {journal} {Phys.
  Rev. A}\ }\textbf {\bibinfo {volume} {77}},\ \bibinfo {pages} {012509}
  (\bibinfo {year} {2008})}\BibitemShut {NoStop}%
\bibitem [{\citenamefont {Perdew}\ \emph {et~al.}(1992)\citenamefont {Perdew},
  \citenamefont {Chevary}, \citenamefont {Vosko}, \citenamefont {Jackson},
  \citenamefont {Pederson}, \citenamefont {Singh},\ and\ \citenamefont
  {Fiolhais}}]{PhysRevB.46.6671}%
  \BibitemOpen
  \bibfield  {author} {\bibinfo {author} {\bibfnamefont {J.~P.}\ \bibnamefont
  {Perdew}}, \bibinfo {author} {\bibfnamefont {J.~A.}\ \bibnamefont {Chevary}},
  \bibinfo {author} {\bibfnamefont {S.~H.}\ \bibnamefont {Vosko}}, \bibinfo
  {author} {\bibfnamefont {K.~A.}\ \bibnamefont {Jackson}}, \bibinfo {author}
  {\bibfnamefont {M.~R.}\ \bibnamefont {Pederson}}, \bibinfo {author}
  {\bibfnamefont {D.~J.}\ \bibnamefont {Singh}}, \ and\ \bibinfo {author}
  {\bibfnamefont {C.}~\bibnamefont {Fiolhais}},\ }\bibfield  {title} {\enquote
  {\bibinfo {title} {Atoms, molecules, solids, and surfaces: Applications of
  the generalized gradient approximation for exchange and correlation},}\
  }\href@noop {} {\bibfield  {journal} {\bibinfo  {journal} {Phys. Rev. B}\
  }\textbf {\bibinfo {volume} {46}},\ \bibinfo {pages} {6671--6687} (\bibinfo
  {year} {1992})}\BibitemShut {NoStop}%
\bibitem [{\citenamefont {Porezag}\ and\ \citenamefont
  {Pederson}(1999)}]{porezag1999optimization}%
  \BibitemOpen
  \bibfield  {author} {\bibinfo {author} {\bibfnamefont {D.}~\bibnamefont
  {Porezag}}\ and\ \bibinfo {author} {\bibfnamefont {M.~R.}\ \bibnamefont
  {Pederson}},\ }\bibfield  {title} {\enquote {\bibinfo {title} {Optimization
  of gaussian basis sets for density-functional calculations},}\ }\href
  {\doibase 10.1103/PhysRevA.60.2840} {\bibfield  {journal} {\bibinfo
  {journal} {Phys. Rev. A}\ }\textbf {\bibinfo {volume} {60}},\ \bibinfo
  {pages} {2840} (\bibinfo {year} {1999})}\BibitemShut {NoStop}%
\bibitem [{\citenamefont {Pederson}\ and\ \citenamefont
  {Jackson}(1990)}]{pederson1990variational}%
  \BibitemOpen
  \bibfield  {author} {\bibinfo {author} {\bibfnamefont {M.~R.}\ \bibnamefont
  {Pederson}}\ and\ \bibinfo {author} {\bibfnamefont {K.~A.}\ \bibnamefont
  {Jackson}},\ }\bibfield  {title} {\enquote {\bibinfo {title} {Variational
  mesh for quantum-mechanical simulations},}\ }\href@noop {} {\bibfield
  {journal} {\bibinfo  {journal} {Phys. Rev. B}\ }\textbf {\bibinfo {volume}
  {41}},\ \bibinfo {pages} {7453--7461} (\bibinfo {year} {1990})}\BibitemShut
  {NoStop}%
\bibitem [{\citenamefont {Zhang}\ and\ \citenamefont
  {Yang}(1998)}]{zhang1998challenge}%
  \BibitemOpen
  \bibfield  {author} {\bibinfo {author} {\bibfnamefont {Y.}~\bibnamefont
  {Zhang}}\ and\ \bibinfo {author} {\bibfnamefont {W.}~\bibnamefont {Yang}},\
  }\bibfield  {title} {\enquote {\bibinfo {title} {A challenge for density
  functionals: Self-interaction error increases for systems with a noninteger
  number of electrons},}\ }\href@noop {} {\bibfield  {journal} {\bibinfo
  {journal} {J. Chem. Phys.}\ }\textbf {\bibinfo {volume} {109}},\ \bibinfo
  {pages} {2604--2608} (\bibinfo {year} {1998})}\BibitemShut {NoStop}%
\bibitem [{\citenamefont {Ruzsinszky}, \citenamefont {Perdew},\ and\
  \citenamefont {Csonka}(2005)}]{ruzsinszky2005binding}%
  \BibitemOpen
  \bibfield  {author} {\bibinfo {author} {\bibfnamefont {A.}~\bibnamefont
  {Ruzsinszky}}, \bibinfo {author} {\bibfnamefont {J.~P.}\ \bibnamefont
  {Perdew}}, \ and\ \bibinfo {author} {\bibfnamefont {G.~I.}\ \bibnamefont
  {Csonka}},\ }\bibfield  {title} {\enquote {\bibinfo {title} {Binding energy
  curves from nonempirical density functionals. {I}. covalent bonds in
  closed-shell and radical molecules},}\ }\href@noop {} {\bibfield  {journal}
  {\bibinfo  {journal} {J. Phys. Chem. A}\ }\textbf {\bibinfo {volume} {109}},\
  \bibinfo {pages} {11006--11014} (\bibinfo {year} {2005})}\BibitemShut
  {NoStop}%
\bibitem [{\citenamefont {Bryenton}\ \emph {et~al.}(2023)\citenamefont
  {Bryenton}, \citenamefont {Adeleke}, \citenamefont {Dale},\ and\
  \citenamefont {Johnson}}]{bryenton2023delocalization}%
  \BibitemOpen
  \bibfield  {author} {\bibinfo {author} {\bibfnamefont {K.~R.}\ \bibnamefont
  {Bryenton}}, \bibinfo {author} {\bibfnamefont {A.~A.}\ \bibnamefont
  {Adeleke}}, \bibinfo {author} {\bibfnamefont {S.~G.}\ \bibnamefont {Dale}}, \
  and\ \bibinfo {author} {\bibfnamefont {E.~R.}\ \bibnamefont {Johnson}},\
  }\bibfield  {title} {\enquote {\bibinfo {title} {Delocalization error: The
  greatest outstanding challenge in density-functional theory},}\ }\href
  {\doibase 10.1002/wcms.1631} {\bibfield  {journal} {\bibinfo  {journal}
  {Wiley Interdiscip. Rev. Comput. Mol. Sci.}\ }\textbf {\bibinfo {volume}
  {13}},\ \bibinfo {pages} {e1631} (\bibinfo {year} {2023})}\BibitemShut
  {NoStop}%
\bibitem [{\citenamefont {Bao}, \citenamefont {Gagliardi},\ and\ \citenamefont
  {Truhlar}(2018)}]{bao2018self}%
  \BibitemOpen
  \bibfield  {author} {\bibinfo {author} {\bibfnamefont {J.~L.}\ \bibnamefont
  {Bao}}, \bibinfo {author} {\bibfnamefont {L.}~\bibnamefont {Gagliardi}}, \
  and\ \bibinfo {author} {\bibfnamefont {D.~G.}\ \bibnamefont {Truhlar}},\
  }\bibfield  {title} {\enquote {\bibinfo {title} {Self-interaction error in
  density functional theory: An appraisal},}\ }\href@noop {} {\bibfield
  {journal} {\bibinfo  {journal} {J. Phys. Chem. Lett.}\ }\textbf {\bibinfo
  {volume} {9}},\ \bibinfo {pages} {2353--2358} (\bibinfo {year}
  {2018})}\BibitemShut {NoStop}%
\bibitem [{\citenamefont {Maier}, \citenamefont {Arbuznikov},\ and\
  \citenamefont {Kaupp}(2019)}]{maier2019local}%
  \BibitemOpen
  \bibfield  {author} {\bibinfo {author} {\bibfnamefont {T.~M.}\ \bibnamefont
  {Maier}}, \bibinfo {author} {\bibfnamefont {A.~V.}\ \bibnamefont
  {Arbuznikov}}, \ and\ \bibinfo {author} {\bibfnamefont {M.}~\bibnamefont
  {Kaupp}},\ }\bibfield  {title} {\enquote {\bibinfo {title} {Local hybrid
  functionals: Theory, implementation, and performance of an emerging new tool
  in quantum chemistry and beyond},}\ }\href@noop {} {\bibfield  {journal}
  {\bibinfo  {journal} {Wiley Interdiscip. Rev. Comput. Mol. Sci.}\ }\textbf
  {\bibinfo {volume} {9}},\ \bibinfo {pages} {e1378} (\bibinfo {year}
  {2019})}\BibitemShut {NoStop}%
\bibitem [{\citenamefont {Vydrov}, \citenamefont {Scuseria},\ and\
  \citenamefont {Perdew}(2007)}]{vydrov2007tests}%
  \BibitemOpen
  \bibfield  {author} {\bibinfo {author} {\bibfnamefont {O.~A.}\ \bibnamefont
  {Vydrov}}, \bibinfo {author} {\bibfnamefont {G.~E.}\ \bibnamefont
  {Scuseria}}, \ and\ \bibinfo {author} {\bibfnamefont {J.~P.}\ \bibnamefont
  {Perdew}},\ }\bibfield  {title} {\enquote {\bibinfo {title} {Tests of
  functionals for systems with fractional electron number},}\ }\href@noop {}
  {\bibfield  {journal} {\bibinfo  {journal} {J. Chem. Phys.}\ }\textbf
  {\bibinfo {volume} {126}} (\bibinfo {year} {2007})}\BibitemShut {NoStop}%
\bibitem [{\citenamefont {Bao}\ \emph {et~al.}(2017)\citenamefont {Bao},
  \citenamefont {Wang}, \citenamefont {He}, \citenamefont {Gagliardi},\ and\
  \citenamefont {Truhlar}}]{doi:10.1021/acs.jpclett.7b02705}%
  \BibitemOpen
  \bibfield  {author} {\bibinfo {author} {\bibfnamefont {J.~L.}\ \bibnamefont
  {Bao}}, \bibinfo {author} {\bibfnamefont {Y.}~\bibnamefont {Wang}}, \bibinfo
  {author} {\bibfnamefont {X.}~\bibnamefont {He}}, \bibinfo {author}
  {\bibfnamefont {L.}~\bibnamefont {Gagliardi}}, \ and\ \bibinfo {author}
  {\bibfnamefont {D.~G.}\ \bibnamefont {Truhlar}},\ }\bibfield  {title}
  {\enquote {\bibinfo {title} {Multiconfiguration pair-density functional
  theory is free from delocalization error},}\ }\href {\doibase
  10.1021/acs.jpclett.7b02705} {\bibfield  {journal} {\bibinfo  {journal} {J.
  Phys. Chem. Lett.}\ }\textbf {\bibinfo {volume} {8}},\ \bibinfo {pages}
  {5616--5620} (\bibinfo {year} {2017})},\ \Eprint
  {http://arxiv.org/abs/https://doi.org/10.1021/acs.jpclett.7b02705}
  {https://doi.org/10.1021/acs.jpclett.7b02705} \BibitemShut {NoStop}%
\bibitem [{\citenamefont {Cohen}, \citenamefont {Mori-S{\'a}nchez},\ and\
  \citenamefont {Yang}(2012)}]{cohen2012challenges}%
  \BibitemOpen
  \bibfield  {author} {\bibinfo {author} {\bibfnamefont {A.~J.}\ \bibnamefont
  {Cohen}}, \bibinfo {author} {\bibfnamefont {P.}~\bibnamefont
  {Mori-S{\'a}nchez}}, \ and\ \bibinfo {author} {\bibfnamefont
  {W.}~\bibnamefont {Yang}},\ }\bibfield  {title} {\enquote {\bibinfo {title}
  {Challenges for density functional theory},}\ }\href@noop {} {\bibfield
  {journal} {\bibinfo  {journal} {Chem. Rev.}\ }\textbf {\bibinfo {volume}
  {112}},\ \bibinfo {pages} {289--320} (\bibinfo {year} {2012})}\BibitemShut
  {NoStop}%
\bibitem [{\citenamefont {Janak}(1978)}]{janak1978proof}%
  \BibitemOpen
  \bibfield  {author} {\bibinfo {author} {\bibfnamefont {J.~F.}\ \bibnamefont
  {Janak}},\ }\bibfield  {title} {\enquote {\bibinfo {title} {Proof that $\frac
  {\partial{}e} {\partial{}n i}$= $\varepsilon$ in density-functional
  theory},}\ }\href {\doibase 10.1103/PhysRevB.18.7165} {\bibfield  {journal}
  {\bibinfo  {journal} {Phys. Rev. B}\ }\textbf {\bibinfo {volume} {18}},\
  \bibinfo {pages} {7165} (\bibinfo {year} {1978})}\BibitemShut {NoStop}%
\bibitem [{\citenamefont {Cohen}, \citenamefont {Mori-S{\'a}nchez},\ and\
  \citenamefont {Yang}(2008)}]{cohen2008fractional}%
  \BibitemOpen
  \bibfield  {author} {\bibinfo {author} {\bibfnamefont {A.~J.}\ \bibnamefont
  {Cohen}}, \bibinfo {author} {\bibfnamefont {P.}~\bibnamefont
  {Mori-S{\'a}nchez}}, \ and\ \bibinfo {author} {\bibfnamefont
  {W.}~\bibnamefont {Yang}},\ }\bibfield  {title} {\enquote {\bibinfo {title}
  {Fractional charge perspective on the band gap in density-functional
  theory},}\ }\href@noop {} {\bibfield  {journal} {\bibinfo  {journal} {Phys.
  Rev. B}\ }\textbf {\bibinfo {volume} {77}},\ \bibinfo {pages} {115123}
  (\bibinfo {year} {2008})}\BibitemShut {NoStop}%
\bibitem [{\citenamefont {Vargas}\ \emph {et~al.}(2020)\citenamefont {Vargas},
  \citenamefont {Ufondu}, \citenamefont {Baruah}, \citenamefont {Yamamoto},
  \citenamefont {Jackson},\ and\ \citenamefont {Zope}}]{vargas2020importance}%
  \BibitemOpen
  \bibfield  {author} {\bibinfo {author} {\bibfnamefont {J.}~\bibnamefont
  {Vargas}}, \bibinfo {author} {\bibfnamefont {P.}~\bibnamefont {Ufondu}},
  \bibinfo {author} {\bibfnamefont {T.}~\bibnamefont {Baruah}}, \bibinfo
  {author} {\bibfnamefont {Y.}~\bibnamefont {Yamamoto}}, \bibinfo {author}
  {\bibfnamefont {K.~A.}\ \bibnamefont {Jackson}}, \ and\ \bibinfo {author}
  {\bibfnamefont {R.~R.}\ \bibnamefont {Zope}},\ }\bibfield  {title} {\enquote
  {\bibinfo {title} {Importance of self-interaction-error removal in density
  functional calculations on water cluster anions},}\ }\href@noop {} {\bibfield
   {journal} {\bibinfo  {journal} {Phys. Chem. Chem. Phys.}\ }\textbf {\bibinfo
  {volume} {22}},\ \bibinfo {pages} {3789--3799} (\bibinfo {year}
  {2020})}\BibitemShut {NoStop}%
\bibitem [{\citenamefont {Ufondu}\ \emph {et~al.}(2023)\citenamefont {Ufondu},
  \citenamefont {Chang}, \citenamefont {Baruah},\ and\ \citenamefont
  {Zope}}]{ufondu2023vertical}%
  \BibitemOpen
  \bibfield  {author} {\bibinfo {author} {\bibfnamefont {P.}~\bibnamefont
  {Ufondu}}, \bibinfo {author} {\bibfnamefont {P.-H.}\ \bibnamefont {Chang}},
  \bibinfo {author} {\bibfnamefont {T.}~\bibnamefont {Baruah}}, \ and\ \bibinfo
  {author} {\bibfnamefont {R.~R.}\ \bibnamefont {Zope}},\ }\bibfield  {title}
  {\enquote {\bibinfo {title} {Vertical detachment energies of ammonia cluster
  anions using self-interaction-corrected methods},}\ }\href@noop {} {\bibfield
   {journal} {\bibinfo  {journal} {J. Chem. Phys.}\ }\textbf {\bibinfo {volume}
  {158}} (\bibinfo {year} {2023})}\BibitemShut {NoStop}%
\bibitem [{\citenamefont {{Johnson III}}(2013)}]{johnson2013nist}%
  \BibitemOpen
  \bibfield  {author} {\bibinfo {author} {\bibfnamefont {R.~D.}\ \bibnamefont
  {{Johnson III}}},\ }\bibfield  {title} {\enquote {\bibinfo {title} {{NIST}
  computational chemistry comparison and benchmark database, {NIST} standard
  reference database number 101},}\ }\href@noop {} {\bibfield  {journal}
  {\bibinfo  {journal} {Release 16a http://cccbdb. nist. gov/(accessed Mar 13,
  2015)}\ } (\bibinfo {year} {2013})}\BibitemShut {NoStop}%
\bibitem [{\citenamefont {Li}, \citenamefont {Requist},\ and\ \citenamefont
  {Gross}(2018)}]{li2018density}%
  \BibitemOpen
  \bibfield  {author} {\bibinfo {author} {\bibfnamefont {C.}~\bibnamefont
  {Li}}, \bibinfo {author} {\bibfnamefont {R.}~\bibnamefont {Requist}}, \ and\
  \bibinfo {author} {\bibfnamefont {E.}~\bibnamefont {Gross}},\ }\bibfield
  {title} {\enquote {\bibinfo {title} {Density functional theory of electron
  transfer beyond the {B}orn-{O}ppenheimer approximation: Case study of
  {LiF}},}\ }\href@noop {} {\bibfield  {journal} {\bibinfo  {journal} {J. Chem.
  Phys.}\ }\textbf {\bibinfo {volume} {148}} (\bibinfo {year}
  {2018})}\BibitemShut {NoStop}%
\bibitem [{\citenamefont {Yamamoto}\ \emph {et~al.}(2019)\citenamefont
  {Yamamoto}, \citenamefont {Diaz}, \citenamefont {Basurto}, \citenamefont
  {Jackson}, \citenamefont {Baruah},\ and\ \citenamefont
  {Zope}}]{yamamoto2019fermi}%
  \BibitemOpen
  \bibfield  {author} {\bibinfo {author} {\bibfnamefont {Y.}~\bibnamefont
  {Yamamoto}}, \bibinfo {author} {\bibfnamefont {C.~M.}\ \bibnamefont {Diaz}},
  \bibinfo {author} {\bibfnamefont {L.}~\bibnamefont {Basurto}}, \bibinfo
  {author} {\bibfnamefont {K.~A.}\ \bibnamefont {Jackson}}, \bibinfo {author}
  {\bibfnamefont {T.}~\bibnamefont {Baruah}}, \ and\ \bibinfo {author}
  {\bibfnamefont {R.~R.}\ \bibnamefont {Zope}},\ }\bibfield  {title} {\enquote
  {\bibinfo {title} {{F}ermi-{L}{\"o}wdin orbital self-interaction correction
  using the strongly constrained and appropriately normed meta-{GGA}
  functional},}\ }\href {\doibase 10.1063/1.5120532} {\bibfield  {journal}
  {\bibinfo  {journal} {J. Chem. Phys.}\ }\textbf {\bibinfo {volume} {151}},\
  \bibinfo {pages} {154105} (\bibinfo {year} {2019})}\BibitemShut {NoStop}%
\bibitem [{\citenamefont {Goerigk}\ \emph {et~al.}(2017)\citenamefont
  {Goerigk}, \citenamefont {Hansen}, \citenamefont {Bauer}, \citenamefont
  {Ehrlich}, \citenamefont {Najibi},\ and\ \citenamefont
  {Grimme}}]{C7CP04913G}%
  \BibitemOpen
  \bibfield  {author} {\bibinfo {author} {\bibfnamefont {L.}~\bibnamefont
  {Goerigk}}, \bibinfo {author} {\bibfnamefont {A.}~\bibnamefont {Hansen}},
  \bibinfo {author} {\bibfnamefont {C.}~\bibnamefont {Bauer}}, \bibinfo
  {author} {\bibfnamefont {S.}~\bibnamefont {Ehrlich}}, \bibinfo {author}
  {\bibfnamefont {A.}~\bibnamefont {Najibi}}, \ and\ \bibinfo {author}
  {\bibfnamefont {S.}~\bibnamefont {Grimme}},\ }\bibfield  {title} {\enquote
  {\bibinfo {title} {A look at the density functional theory zoo with the
  advanced {GMTKN55} database for general main group thermochemistry{,}
  kinetics and noncovalent interactions},}\ }\href@noop {} {\bibfield
  {journal} {\bibinfo  {journal} {Phys. Chem. Chem. Phys.}\ }\textbf {\bibinfo
  {volume} {19}},\ \bibinfo {pages} {32184--32215} (\bibinfo {year}
  {2017})}\BibitemShut {NoStop}%
\bibitem [{\citenamefont {Vydrov}\ and\ \citenamefont
  {Scuseria}(2004)}]{vydrov2004effect}%
  \BibitemOpen
  \bibfield  {author} {\bibinfo {author} {\bibfnamefont {O.~A.}\ \bibnamefont
  {Vydrov}}\ and\ \bibinfo {author} {\bibfnamefont {G.~E.}\ \bibnamefont
  {Scuseria}},\ }\bibfield  {title} {\enquote {\bibinfo {title} {Effect of the
  {P}erdew--{Z}unger self-interaction correction on the thermochemical
  performance of approximate density functionals},}\ }\href@noop {} {\bibfield
  {journal} {\bibinfo  {journal} {J. Chem. Phys.}\ }\textbf {\bibinfo {volume}
  {121}},\ \bibinfo {pages} {8187--8193} (\bibinfo {year} {2004})}\BibitemShut
  {NoStop}%
\bibitem [{\citenamefont {Aquino}, \citenamefont {Shinde},\ and\ \citenamefont
  {Wong}(2020)}]{https://doi.org/10.1002/jcc.26168}%
  \BibitemOpen
  \bibfield  {author} {\bibinfo {author} {\bibfnamefont {F.~W.}\ \bibnamefont
  {Aquino}}, \bibinfo {author} {\bibfnamefont {R.}~\bibnamefont {Shinde}}, \
  and\ \bibinfo {author} {\bibfnamefont {B.~M.}\ \bibnamefont {Wong}},\
  }\bibfield  {title} {\enquote {\bibinfo {title} {Fractional occupation
  numbers and self-interaction correction-scaling methods with the
  {F}ermi-{L}{\"o}wdin orbital self-interaction correction approach},}\
  }\href@noop {} {\bibfield  {journal} {\bibinfo  {journal} {J. of Comput.
  Chem.}\ }\textbf {\bibinfo {volume} {41}},\ \bibinfo {pages} {1200--1208}
  (\bibinfo {year} {2020})}\BibitemShut {NoStop}%
\end{thebibliography}%

\end{document}